\documentclass[3p,authoryear,semicolon]{article}

\usepackage[english]{babel}
\usepackage[utf8]{inputenc}

\usepackage{graphicx}
\usepackage{amsmath}
\usepackage{amssymb}
\usepackage{geometry}
\usepackage{booktabs}
\usepackage{upgreek}
\usepackage{adjustbox}

\usepackage{pdfpages}

\usepackage{natbib}
\usepackage{multicol}
\usepackage{multirow}

\usepackage[hidelinks]{hyperref}
\usepackage{lineno}

\usepackage{caption}
\usepackage{subcaption}
\usepackage{float}

\DeclareMathOperator*{\argmax}{arg\,max}
\DeclareMathOperator*{\argmin}{arg\,min}

\title{Temporal segmentation of motion propagation in response to an external impulse}

\author{Sina Feldmann$^{1,2}$, Thomas Chatagnon$^3$, Juliane Adrian$^{1,*}$, Julien Pettré$^3$, Armin Seyfried$^{1,2}$}

\date{
	\small{\textit{
		$^1$ Institute for Advanced Simulation 7: Civil Safety Research, Forschungszentrum J\"ulich, J\"ulich, Germany\\%
		$^2$ Faculty of Architecture and Civil Engineering, University of Wuppertal, Wuppertal, Germany\\%
		$^3$ Univ Rennes, Inria, CNRS, IRISA, M2S, France\\%
		$^*$ Correspoding author, email: j.adrian@fz-juelich.de
	}}	
}

\begin{document}

\maketitle

\begin{abstract}
In high-density crowds, local motion can propagate, amplify, and lead to macroscopic phenomena, including "density waves". These density waves only occur when individuals interact, and impulses are transferred to neighbours. How this impulse is passed on by the human body and which effects this has on individuals is still not fully understood.
To further investigate this, experiments focusing on the propagation of a push were conducted within the EU-funded project CrowdDNA. 
In the experiments the crowd is greatly simplified by five people lining up in a row. The rearmost person in the row was pushed forward in a controlled manner with a punching bag. The intensity of the push, the initial distance between participants and the initial arm posture were varied. Collected data included side view and top view video recordings, head trajectories, 3D motion using motion capturing (MoCap) suits as well as pressure measured at the punching bag. With a hybrid tracking algorithm, the MoCap data are combined with the head trajectories to allow an analysis of the motion of each limb in relation to other persons.

The observed motion of the body in response to the push can be divided into three phases. These are (i) receiving an impulse, (ii) receiving and passing on an impulse, and (iii) passing on an impulse. 
Using the 3D MoCap data, we can identify the start and end times of each phase.
To determine when a push is passed on, the forward motion of the person in front has to be considered. 
The projection of the center of mass relative to the initial position of the feet is a measure of the extent to which a person is displaced from the rest position. 
Specifying the timing of these phases is particularly important to distinguish between different types of physical interactions.
Our results contribute to the development and validation of a pedestrian model for identifying risks due to motion propagation in dense crowds.

\end{abstract}


\paragraph{Ethical Review:} 
The experiments were approved by the ethics board of the University of Wuppertal in April 2021 (Reference: \mbox{MS/BBL 210409 Seyfried}).

\paragraph{Data Accessibility:}
The raw data from the experiments are freely accessible at the Pedestrian Dynamics Data Archive of the Research Centre J\"ulich, https://doi.org/10.34735/ped.2022.2. The data include videos in mp4 format, head trajectories as txt files, combined MoCap data as c3d files, pressure data as txt files and information about participants.

\paragraph{Funding:}
This study was funded by the European Unions Horizon 2020 research and innovation program within the project CrowdDNA [grant number 899739].
SF was additionally supported by a scholarship from the German Academic Exchange Service (DAAD).

\paragraph{Acknowledgements:}
We would like to thank Jernej {\v C}amernik, Marc Ernst, Helena L\"ugering and Anna Sieben who supported us in the conception, planning and realisation of the experiments.
Furthermore, we are grateful to Alica Kandler, Maik Boltes, Ann Katrin Boomers and Tobias Schr\"odter for their help in setting up the experiments and curating the data.

\section{Introduction} 

Motion is a fundamental concept in physics and refers to the change in position of an object, either in terms of its distance from a point or its displacement from an initial position. Studying the motion of pedestrians is relevant for applications such as public space design, traffic engineering, and crowd safety. 
Especially in large crowds, pedestrian safety is of great importance, and crowd accidents are widely reported in the media (\cite{feliciani_trends_2023}). 
To better understand the causes of these accidents and the associated risks, it is necessary to analyse crowd motions.\\

The motion of a crowd (macroscopic scale) results from the motion of multiple individuals and, in particular, from the motion of their three-dimensional bodies (submicroscopic scale).
Macroscopic phenomena are caused by the propagation of motion in a crowd, which occur when individuals interact and impulses are transmitted.
The motion of people in crowds is generally described by points in terms of their velocity or flow through a facility or characterised by their density distributions (\cite{ meyers_empirical_2018, feng_data_2021}).
In this context, the motion of pedestrians is assessed in two dimensions, either when trajectories are recorded in experiments (e.g. \cite{adrian_crowds_2020}) or when pedestrians are represented as circles or ellipses in models (\cite{meyers_modelling_2018, korbmacher_review_2022}). 
Videos of crowds are also evaluated during on-site observations (\cite{bottinelli_can_2018, sundararaman_tracking_2021}) or when analysing actual crowd accidents (e.g. \cite{helbing_dynamics_2007, jiayue_comparison_2014}), considering mainly the motion of heads as the main body is occluded.
For a detailed list of accidents that occurred over the years see \cite{feliciani_trends_2023}.
In this way, macroscopic phenomena and collective movements of a crowd that are considered high risk can be characterised. 
These include, for example, clogging in front of bottlenecks (\cite{muir_effects_1996, garcimartin_flow_2016}), density waves (\cite{bottinelli_emergent_2016}), pressure waves (\cite{feliciani_systematic_2020, sieben_inside_2023}), turbulences (\cite{helbing_dynamics_2007, krausz_loveparade_2012}), or transversal waves (\cite{garcimartin_redefining_2018, adrian_crowds_2020}).  
There are approaches to simulate contact forces (\cite{kim_velocity-based_2015, van_toll_sph_2021}), but how exactly impulses propagate in a crowd is not yet fully understood.
Furthermore, the exact motion of people is disregarded and three dimensional movement patterns cannot be described sufficiently.
One example of this is the risk of people falling over and forming a pile on the ground (\cite{helbing_crowd_2012, sieben_inside_2023}).\\

The risk of losing balance, stumbling or falling down results from challenges that individuals face at the sub-microscopic scale.
Previous research on human standing balance has focused on different recovery strategies for single individuals following external perturbations, e.g. the ankle, hip or stepping strategies to regain balance (\cite{winter_human_1995, maki_control_2006, tokur_review_2020}).
In this context, the critical point for static human balance has been defined as the time at which the projected centre of mass (CoM) on the ground passes the base of Support (BoS) in static conditions (\cite{winter_human_1995}). Later on, the use of the extrapolated center of mass (XCoM) was proposed to study standing balance in dynamic conditions (\cite{hof_condition_2005, hof_extrapolated_2008}).
Further indicators for human standing balance such as the Margin of Stability (MoS) (\cite{hof_condition_2005, rosenblatt_measures_2010, hak_steps_2013}) or the Time-to-boundary (\cite{schulz_can_2006, emmens_predicting_2020,chatagnon_stepping_2023}), have also been introduced.
The aforementioned studies are limited to single individuals and do not consider how the presence of others restricts the torso and limb movements, thus affecting the recovery of balance.
On the one hand, the lack of space may alter the strategies chosen and foster stumbling; on the other hand, individuals may also use other change-in-support options, such as contact with their neighbours, to increase their BoS (\cite{maki_1997}).
Both possibilities were mentioned in witness statements at the Loveparade (\cite{sieben_inside_2023}).\\

First experimental studies in which people were pushed and unbalanced in a crowd have been conducted only on small groups.
They mainly examine the collision dynamics (\cite{wang_modeling_2019}), contact forces (\cite{li_experimental_2020}) and pressure measurements (\cite{wang_study_2018}) in groups of people or the motion of only two participants (\cite{li_experimental_2021}).
This does not resolve the difficulty of understanding individual three-dimensional motion, the way people regain balance and how this motion propagates in a crowd. 
Furthermore, \cite{sieben_inside_2023} concluded that both, falls and macroscopic waves, cause the greatest risk in dense crowds.
Therefore, it is essential to connect the macroscopic with the sub-microscopic analysis, when studying the motion of pedestrians.\\

With our research, we aim to develop a methodology to comprehend the dynamics of a human body when losing balance while the movement of the torso and limbs are restricted by surrounding people.
Therefore, we conducted a series of experiments to investigate the propagation of a push and the balance recovery in a standing crowd. 
This experimental campaign was carried out as part of the EU-funded project CrowdDNA (\cite{CrowdDNAProject2022}). 
For a starting point of this research direction, we simplified the crowd to a row of five people and varied specific parameters, such as the intensity of the impulses and the initial inter-person distance in order to study their effect on the impulse propagation (\cite{feldmann_forward_2023}).
We collected head trajectories and 3D motion capturing data of each individual, which we then combined to one data-set to get an insight into how the individuals interact with each other by describing the movement of their body in three dimensions.
In a first analysis of the macroscopic characteristics, the propagation of pushes in terms of distance and speed were investigated in one dimension (the direction of the push) without considering the individual limb motions of the participants.
Now we take the analysis a step further and examine the same experiments in more detail for the 3D reactions of each participant.\\

To gain a deeper understanding into how impulses propagate through a row of people and how this is influenced by individual behaviour, we have to consider not only reactions to the external impulse, but also the interactions between people.
In this context, the term "impulse" refers to the transfer of a force over a period of time causing an external perturbation to the participants.
Dividing what happens into different phases facilitates a detailed analysis in the future, especially to distinguish between various types of physical interactions that occur in each phase.
This allows us to better compare differences between single person trials (as conducted by \cite{chatagnon_stepping_2023}) with the row experiments, to separate between reactions of unintended displacement from the resting position caused by external impulses and the intended balance recovery strategy, as well as to investigate whether an impulse is reduced or intensified along the row. 
By studying the propagation of motion, insights into how pedestrians behave in response to various forces and conditions can be obtained.
This understanding allows pedestrian behaviour to be more accurately described and helps to develop three-dimensional models of pedestrian dynamics also in high density scenarios, in order to hopefully predict and identify potentially dangerous situations in the future.\\

\section{Methods}
\subsection{Experiments}
To investigate motion propagation in crowds, we are analysing the same 97 trials from the pushing experiments as described in \cite{feldmann_forward_2023}.
External impulses were delivered by one of the experimenters manually pushing a punching bag, that was suspended horizontally from the ceiling, towards the upper back of the last participant in a line of five people. All participants standing with feet hip-width apart and facing in the direction of the push.

In the experiment, certain factors were manipulated, including the intensity of the impulses. 
The inter-person distance, measured by individual arm length, was set as arm, elbow or as close as possible (none).
At the beginning of a trial, participants held their arms either up, down, or in an arm position of their choice.
The first person of the queue either had enough space to move forward freely or was standing in front of a wall with the specified distance.

The experiments were recorded from the side to enable a visual qualitative analysis as well as from above to collect individual head trajectories with the help of PeTrack (\cite{BOLTES2013127, boltes_maik_2021_5126562}).
In addition, each participant wore a Xsens (\cite{schepers_xsens_2018}) motion capture (MoCap) suit containing 17 inertial measurement units.
This allowed us to record individual 3D motion data, and then integrate them into a common reference system by fusing the 3D data with the head trajectories (\cite{boltes_hybrid_2021}).
This 3D MoCap data is in the c3d format and contains the positions of 64 points, 22 joints, and the calculated center of mass (CoM) of the human skeleton.
Furthermore, we used a pressure sensor from Xsensor (\cite{XsensorLX210:50.50.05}) to measure the strength of the impulses at the punching bag and a sensor from TekScan (\cite{TekscanPMS5400N}) to estimate the impact of the foremost person on the wall.

\begin{figure}[H]
\centering
\includegraphics[width=\textwidth]{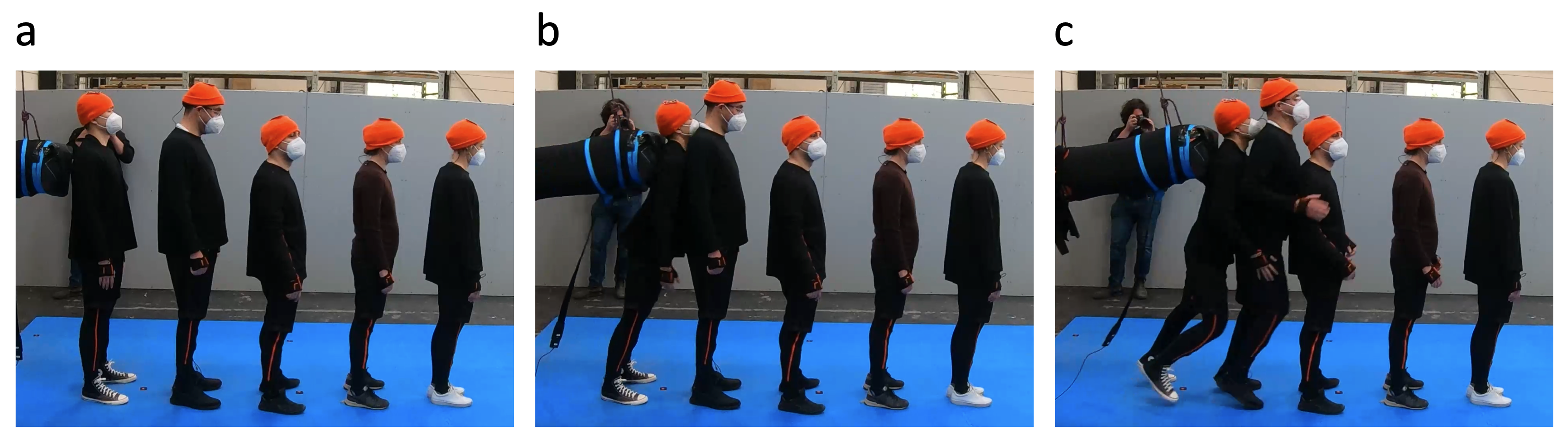}
\caption{Snapshots of the experiments from the side-view camera. Five people lined up along the y-axis in front of a punching bag. (a) At the beginning of the trial, all participants stand at rest. (b) The last person in the row is pushed forward by the punching bag and touches the next person. (c) The impulse reaches the third person of the row.} 
\label{fig:sideview}
\end{figure}

For the analysis, we went for a two method approach: First the side-view videos were qualitatively described and three temporal phases observed.
As a second step, the 3D MoCap data were analysed to define the start and end points of each phase.
We will focus our analysis only on specific points of the human skeleton, which can be found in the Supplementary Information.

\subsection{Analytical methods for the analysis of MoCap data}

For the analysis of the MoCap data, various analytical methods were applied.

\subsubsection{Distance between a point and a line in a 2D plane}

To describe the loss of standing balance, the distance between the projected CoM on the ground, represented as a trajectory, and a line passing through the two toe points is used. 
In three dimensions, the distance between a point and a line can be calculated using the cross product.
The projection on a 2D plane is simplified as follows.

Let $\mathbf{a}, \mathbf{b}, \mathbf{c} \in \mathbb{R}^2 $ and the line, that passes through the points $\mathbf{a}, \mathbf{b}$, is defined as 
$\mathbf{ab} = \mathbf{a} + t \cdot \mathbf{n} $ with unit vector $\mathbf{n} = \frac{(\mathbf{b} - \mathbf{a})}{ \| (\mathbf{b} - \mathbf{a})\|}$ along the line and $t \in \mathbb{R}$.
Then the distance $d(\mathbf{ab}, \mathbf{c})$ between point $\mathbf{c}$ and line $\mathbf{ab}$ is given by:
\begin{equation} 
d(\mathbf{ab}, \mathbf{c}) = n_{1} \cdot  (a_{2} - c_2)  - n_{2} \cdot (a_{1}-c_1) 
\label{eq:distance3}
\end{equation}

\subsubsection{Distance between a point and a line segment in a 2D plane}

The point - line segment distance is used to estimate the distance between hands and the back of the person in front.
A line segment is a special case of a line whereby the segment is limited by two specified endpoints and includes every point on the line that lies between its endpoints.
The distance between a point and the segment is defined as the length of the shortest connecting line between that point and a point on the segment.

Let $\mathbf{a}, \mathbf{b}, \mathbf{c} \in \mathbb{R}^2 $ as discussed above. 
The line segment $\overline{\mathbf{ab}}$ with its two endpoints $\mathbf{a}$ and $\mathbf{b}$ is defined similarly to a line $\overline{\mathbf{ab} }= \mathbf{a} + \tilde{t} \cdot \mathbf{n} $.
Here, $\mathbf{n} $ is again the unit vector $\mathbf{n} = \frac{(\mathbf{b} - \mathbf{a})}{ \| (\mathbf{b} - \mathbf{a})\|}$, but this time $\tilde{t}  \in [0,  \| (\mathbf{b} - \mathbf{a})\|]$.
The distance of point $\mathbf{c}$ to the line segment can be calculated using equation \ref{eq:distance3}:

\begin{equation} 
    d(\overline{\mathbf{ab}}, \mathbf{c}) =
\begin{cases}
    d(\mathbf{ab}, \mathbf{c})  & \text{if } 0 \leq (\mathbf{c} - \mathbf{a}) \cdot \mathbf{n} \leq  \| (\mathbf{b} - \mathbf{a})\|\\
    \min (\| (\mathbf{c} - \mathbf{a})  \|, \| (\mathbf{c} - \mathbf{b})  \|), & \text{otherwise}
\end{cases}
\label{eq:distance}
\end{equation}

\subsubsection{Distance between two line segments in a 2D plane}

To investigate the distance between the chest of a participant and the back of the next person, the distance between two line segments is used.
The distance of two line segments $\overline{\mathbf{ab}}$ and $\overline{\mathbf{cd}}$ is considered as the shortest connecting line between an endpoint of one segment to a point of the other segment.
Therefore, equation \ref{eq:distance} can be used for each endpoint of the segments to determine $d(\overline{\mathbf{ab}}, \overline{\mathbf{cd}})$:

\begin{equation} 
d(\overline{\mathbf{ab}}, \overline{\mathbf{cd}}) = \min[d(\overline{\mathbf{ab}}, \mathbf{c}), d(\overline{\mathbf{ab}}, \mathbf{d}), d(\overline{\mathbf{cd}}, \mathbf{a}), d(\overline{\mathbf{cd}}, \mathbf{b})]
\label{eq:distance2}
\end{equation}

\subsubsection{Forward speed and acceleration}

Velocities $\mathbf{v}_i^p(t)$ and accelerations $\mathbf{a}_i^p(t)$ can be calculated from the position data of individual body parts $\mathbf{r}_i^p(t)$ as follows:

\begin{equation} 
\mathbf{v}_i^p(t) = \frac{\mathbf{r}_i^p(t+\Delta t) - \mathbf{r}_i^p(t-\Delta t)}{2 \cdot \Delta t}
\label{eq:v}
\end{equation}

\begin{equation} 
\mathbf{a}_i^	p(t) = \frac{\mathbf{v}_i^p(t+\Delta t) - \mathbf{v}_i^p(t-\Delta t)}{2 \cdot \Delta t}
\label{eq:a}
\end{equation}

Hereby, $i$ indicates the number of the person in the row ($i \in [1,5]$) and $p$ is a point corresponding to an anatomical landmark of the human body reconstructed from the MoCap data.
We used $\Delta t = 0.05 \,\text{s}$ to better detect rapid changes in the motion of participants, because the duration of the external impulses were quite short with 0.69\,s on average. 
This time step is also shorter than the typical reaction time of young adults, i.e. from $0.18\,\text{s}$ to $0.3\,\text{s}$ for reactions to haptic or visual stimuli (\cite{Peon-2013,Aditya-2015}).
Furthermore, we want to investigate mainly the forward motion of participants which is the same as the pushing direction, because we are particularly interested in the effect of the impulse on the motion.
To determine the forward component, we calculate the unit vector $ \mathbf{u}_i$ in the xy-plane that is perpendicular to the hip vector at the beginning of the trial ($\mathbf{r}_i^{\text{LHIP}}(0) - \mathbf{r}_i^{\text{RHIP}}(0)$).
Then the forward speed $v_i^p(t)$ results from:

\begin{equation} 
  v_i^p(t) = \mathbf{v}_i^p(t) \cdot \mathbf{u}_i
\label{eq:vf}
\end{equation}
 The same applies for the acceleration $\mathbf{a}_i^p(t) $.

 \subsubsection{Projection of 3D position to a plane}

The projection of a 3D vector corresponds to a Matrix-Vector multiplication with the projection matrices 
$$ P_{xy}=
\begin{pmatrix}
1 & 0 & 0 \\
0 & 1 & 0 \\
0 & 0 & 0 \\
\end{pmatrix}, 
 P_{yz}=
\begin{pmatrix}
0 & 0 & 0 \\
0 & 1 & 0 \\
0 & 0 & 1 \\
\end{pmatrix}, 
 P_{xz}=
\begin{pmatrix}
1 & 0 & 0 \\
0 & 0 & 0 \\
0 & 0 & 1 \\
\end{pmatrix}$$

for the xy, yz and xz plane respectively.

\section{Analysis}

\subsection{Qualitative description of phases of motion}

In the experiments, the body of a participant receives an external impulse and the person is displaced from their rest position as a result of the push.
Muscles and skeleton can dampen this impulse, direct it into the ground or transfer it to the person in front.
To manage the impact and to regain a stable position, participants can choose different strategies, for instance, transferring forces into the ground by taking a step, transferring forces through arms and hands to the person in front with and without a step, or dampening the impulse by moving the hips backwards.
These are just a few examples, and due to the complexity of the human body, there are even more possible reactions.
Consequently, various physical interactions can occur along the row.
In order to study impulse propagation in more detail and detect these strategies, it is important to first separate and divide the motions and interactions into different temporal phases.\\

\begin{figure}[H]
\centering
\includegraphics[width=0.9\textwidth]{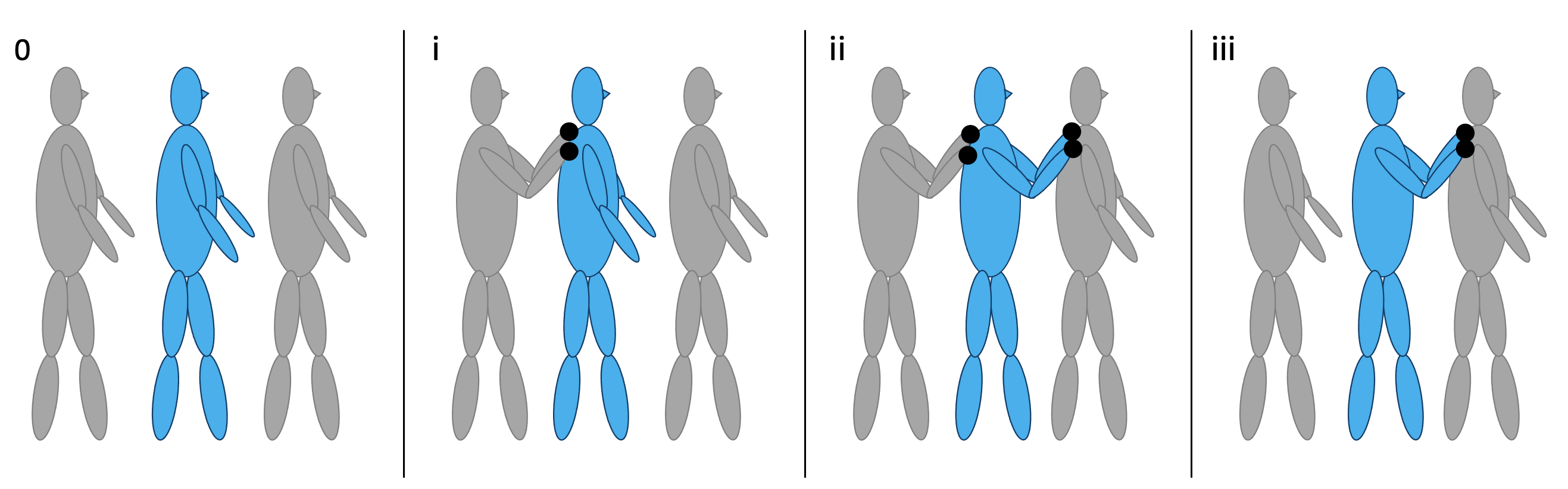}
\caption{Overview of the phases of impulse propagation for the blue person in the middle: (0) no external impact (i) receiving an impulse, (ii) receiving and passing on an impulse, and (iii) passing on an impulse. The black dots represent a contact.}
\label{fig:phases}
\end{figure}

For each person within the row, multiple phases of impulse propagation can be identified, although the number of applicable phases may vary.
In the following, the most detailed version, which consists of three phases, is explained.
These phases are (0) initial position at rest, (i) receiving an impulse, (ii) receiving and passing on an impulse, and (iii) passing on an impulse (Figure \ref{fig:phases}). 
From an observational analysis of the side-view videos, we obtained different individual behaviours within the three phases. 
In this context, touching refers to any contact that may occur between two people, for example, with hands or with the upper body. 
In the first phase, the blue person in the middle of the row is touched from behind, i.e. they receive the impulse, and are consequently displaced from the rest position which can lead to losing balance.
The second phase differs from the first phase by the fact that the person now additionally touches the next person in the row and therefore passes on the impulse.
When the person is no longer touched from behind, the third phase begins. They are not receiving the push anymore, but are still passing it on. This phase lasts until the person has regained balance or does not touch the person in front anymore. 

\subsection{3D data from MoCap}

The second analysis aims to identify characteristics in the 3D MoCap data that represent the different phases in order to accurately determine the start and end times of each phase.
Following the reconstruction from the MoCap data, only certain points corresponding to anatomical landmarks of the human skeleton will be examined.
The points and the abbreviations used as well as the index to find these points in the reconstructed MoCap data are listed in the Supplementary Information.
The analysis was performed in python with the \textit{ezc3d} library (\cite{michaud_ezc3d_2021}).
To detect the phases, several variables, namely the forward velocity, the margin of stability and the distance between two participants, have to be investigated in combination.
In this section, each variable is first explained separately, and then the phases are defined based on the combination.
For temporal segmentation of the phases, we examine several times to derive the actual start and end times of the phases.
A list of the studied times is given in Table \ref{tab:times} as an overview and will be explained in more detail in the following section.

\begin{table}[H]
\centering
    \caption{Overview of the investigated times.}
    \label{tab:times}
    \begin{tabular}{l|lll|l}
    	\toprule
        Description & Variable &  Time  & Condition & Derived time\\
         \midrule
	
	\multirow{ 5}{*}{Start of motion} &    $a_i^{\text{CoM}}(t_1) > 0.3\, \frac{\text{m}}{\text{s}^2} $ & $t_1$&  $< t_2$& \multirow{ 5}{*}{$t_{\text{start}}[i]$ }\\
   				& $v_i^{\text{CoM}}(t_2)> 0.05\, \frac{\text{m}}{\text{s}}$ & $t_2$ & $> t_1$&\\
   				& $v_i^{\text{C7}}(t_2) > 0.05\, \frac{\text{m}}{\text{s}}$ & $t_2$ &  $ > t_1$&\\
  				& $v_i^{\text{HIP}}(t_2) > 0.05\, \frac{\text{m}}{\text{s}}$ & $t_2$ & $ > t_1$& \\
				&   $a_i^{\text{CoM}}(t_0) > 0.15\, \frac{\text{m}}{\text{s}^2} $ & $t_0$& $< t_2$&  \\

	&&&& \\
	 Unstable position & $\argmin_t MoS_i (t)$ & $t_{\text{min}}$ & / & $t_{\text{min}}[i]$ \\
	 &&&& \\
	 \multirow{2}{*}{Stabilised position} &  $MoS_i (t_2)  \geq 0.87 \cdot MoS_i (0)$ & $\min[t_2]$ & $ >  t_{\text{min}}[i]$ &  \multirow{2}{*}{$t_{\text{stable}}[i]$} \\
	 			& $\argmax_{t_3} MoS_i(t_3)$ & $t_3$ & $ > t_{\text{min}}[i] $ & \\
				& & & $< t_{\text{min}}[i]  + 2\,\text{s}$ &  \\
&&&& \\
	End of contact & $d_{\text{min}}[i, i+1](t_4) > 0.12\,\text{m}$ & $\min[t_4]$ & $ > t_{\text{start}}[i+1]$ & $ t_{\text{touch}}[i] $\\
	&&&& \\
	End of phases& $\min[ t_{\text{stable}}[i], t_{\text{touch}}[i] ] $ & /& / &   $t_{\text{end}}[i]$ \\
    \end{tabular}
\end{table}

\subsubsection{Contact}
\label{sec:touching}

As observed in the videos, whether there is a contact or not is a measure of receiving and passing on the impulse.
This can be analysed by looking at the distance between different points of the individual MoCap data. 
In most cases, people are touched at the back either by hands or the upper body.
Contacts that were observed in the videos include hands on upper back, hands on lower back, chest on upper back, belly on lower back and hands on the side of the arms.
It is also possible for multiple contacts to occur at the same time, for example the chest on the upper back and hands on the lower back, or the belly on the lower back and hands on the upper back.

\begin{figure}[H]
 \centering
\includegraphics[width=\textwidth]{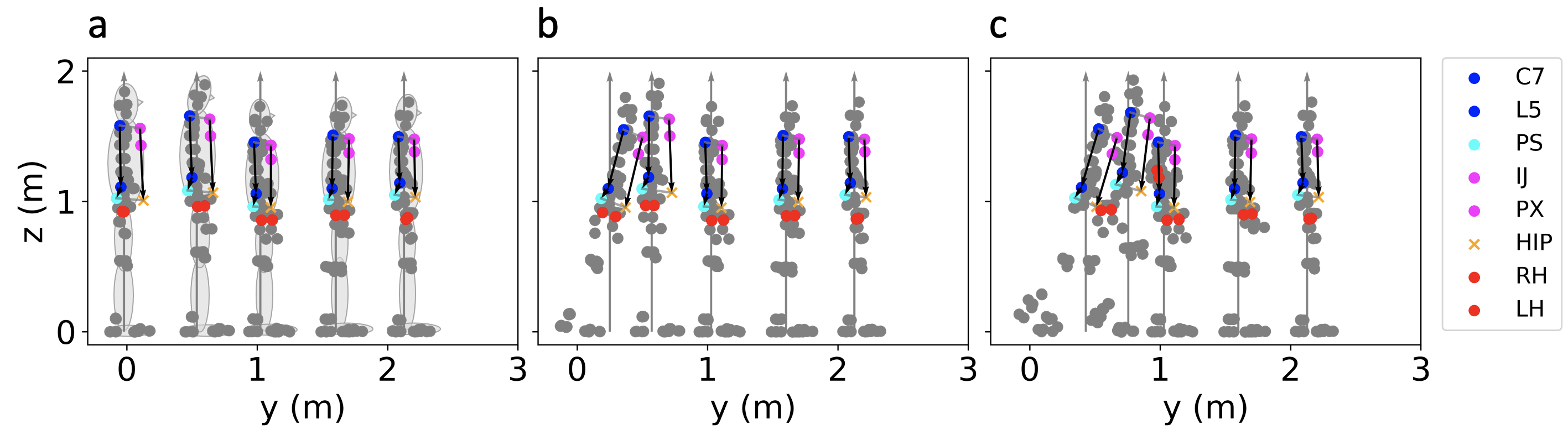}
\caption{MoCap data for the same experiment and times as in Figure \ref{fig:sideview}. The points of the MoCap data for all five participants are shown in grey. The upper body of each participant is approximated by the two points of the sternum IJ and PX (magenta), the height of the hip HIP (yellow), the sacrum PS (cyan) and two points of the spine C7 and L5 (blue). The hands RH and LH are shown as red points and the segments of the chest, upper back and lower back are represented as black arrows. (a) All participants stand at rest. (b) The impulse is passed on to the second person. (c) The second person in the row receives and passes on the impulse.}
\label{fig:upperbody}
\end{figure}

In order to find contacts, we estimate the extent of the torso using the 3D MoCap data, which corresponds to the approximate positions of anatomical landmarks on the human skeleton.
The upper body of each participant is therefore approximated by the two points of the sternum (IJ, PX), the height of the hip (HIP), the sacrum (PS) and two points of the spine C7 and L5 as plotted in Figure \ref{fig:upperbody}.
This results in three line segments: the segment of the chest (CH) from the upper sternum through the lower sternum to the projected hip point, the upper back (UB) from C7 to L5 and the lower back (LB) from L5 to the sacrum. 
For the hands, the points of the right as well as the left palm are considered (RH, LH).

For simplification and because all participants are lined up along the y-axis of the experimental area, the x-component of the MoCap data was not considered. 
To calculate the shortest distance $d_{\text{min}}[i, i+1]$ between the two participants $i$ and $i+1$ with $i\in [1,4]$, equations \ref{eq:distance} and  \ref{eq:distance2} are used for the segments in the yz plane as follows:

\begin{equation*} 
\begin{split}
d_{\text{min}}[i, i+1] = \min[&d(P_{yz}\mathbf{r}_i^{\text{CH}}, P_{yz}\mathbf{r}_{i+1}^{\text{UB}}), d(P_{yz}\mathbf{r}_i^{\text{CH}}, P_{yz}\mathbf{r}_{i+1}^{\text{LB}}), \\ 
&d(P_{yz}\mathbf{r}_i^{\text{RH}}, P_{yz}\mathbf{r}_{i+1}^{\text{UB}}), d(P_{yz}\mathbf{r}_i^{\text{RH}}, P_{yz}\mathbf{r}_{i+1}^{\text{LB}}), \\
&d(P_{yz}\mathbf{r}_i^{\text{LH}}, P_{yz}\mathbf{r}_{i+1}^{\text{UB}}), d(P_{yz}\mathbf{r}_i^{\text{LH}}, P_{yz}\mathbf{r}_{i+1}^{\text{LB}})
]
\end{split}
\end{equation*}

\begin{figure}[H]
 \centering
\includegraphics[width=0.4\textwidth]{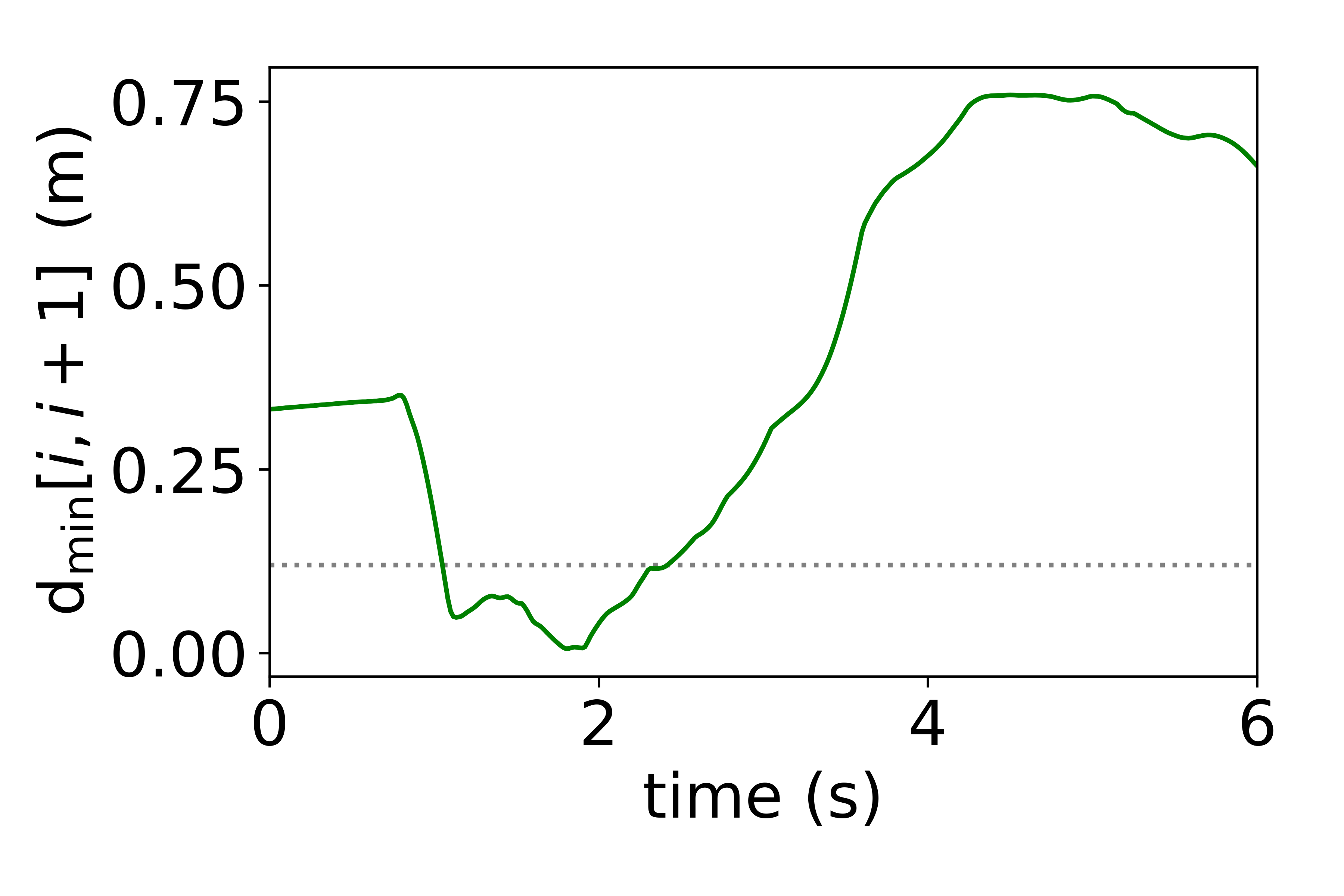}
\caption{An example of the closest distance $d_{\text{min}}[i, i+1] $ changing over time. As a limit for touching, the threshold value of 0.12\,m is represented as dotted line. }
\label{fig:d_time}
\end{figure}

For the shortest inter-person distance to be as accurate as possible, the correct positioning in the experimental area is crucial. 
In addition, to an error of the trajectory data of 1.54\,cm and the 3D MoCap data of approximately 3\,cm, an error resulting from the fusion of both data-sets has to be considered.
The simplified representation of a person by three segments further decreases the accuracy.
Therefore, we define a touch if the distance falls below a certain threshold, which has been set to 0.12\,m by an initial estimation (see Supplementary Information).
An example of the closest distance between two people changing over time is shown in Figure \ref{fig:d_time}.
According to our definition, contact occurs when the distance drops below the dotted line, which can be confirmed by the video.
It can be seen that the two people approach each other and touch between 1\,s and 2.4\,s of the trial.

\begin{figure}[H]
 \centering
\includegraphics[width=0.7\textwidth]{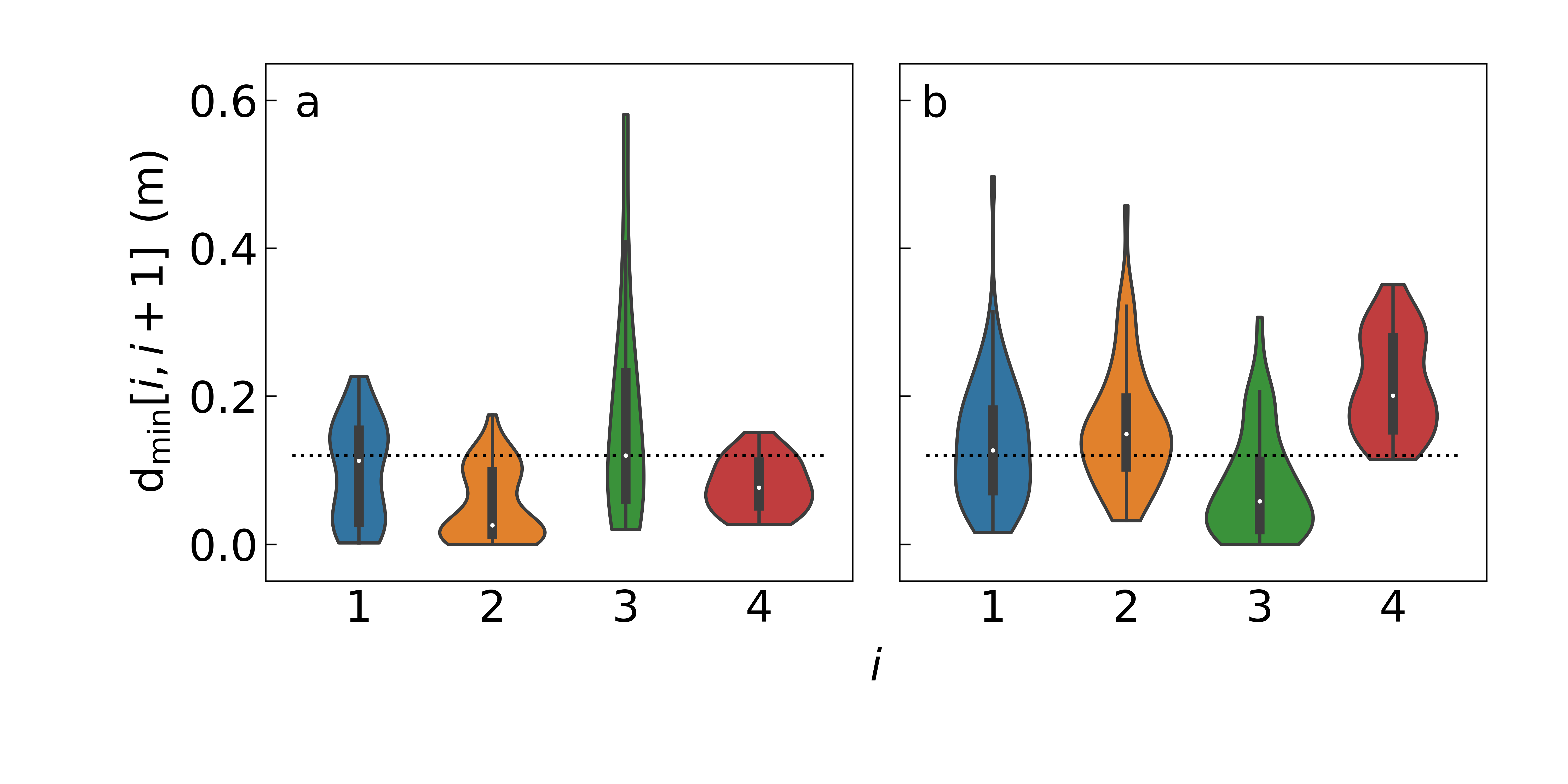}
\caption{Closest distance between two persons $i$ and $i+1$ at the time the motion starts. (a) Series of experiments without a wall and (b) with wall. Within a series, the order of participants remained the same. The person who receives the push first has number one. The dotted line represents a threshold value of 0.12\,m}
\label{fig:touching}
\end{figure}

Furthermore, individual differences might be relevant, as the MoCap data only gives a reconstruction of the skeleton meaning that the exact body dimension cannot be taken into account when calculating the distance. 
To find appropriate thresholds, we assume that a person is touched from behind when they start moving forward, and therefore we analyse the closest distance at the time the motion starts (Figure \ref{fig:touching}).
For the definition of the start of motion see Section \ref{sec:motion}.

Figure \ref{fig:touching} illustrates that the distances between the same participants vary considerably across trials and there is no significant difference between individuals.
Our conclusions are this: The calculated distances are too imprecise for a definition of a phase and touching cannot be reliably detected in the MoCap data.
Nevertheless, the distance can be used subsequently to describe whether people are moving towards or away from each other and to find an overall minimal distance.

\subsubsection{Start of motion}
\label{sec:motion}
In order to define the phases, the start time of the motion is considered for further analysis and the overall definition of the phases has to be adjusted.
The aim of this section is to find the time at which a person starts moving forward because of the external impulse and thus determine the start time of the first phase.
The analysis of the 3D motion of a human body is complex, because the body consists of several limbs that can move individually.
Furthermore, participants can react differently when they are pushed. 
To investigate the motion propagation due to an external impulse more uniformly, the motion of the center of mass (CoM) can be compared.
However, the CoM will also move when, for example, only the arms are raised upwards and the person does not receive the push.
Participants can in addition move independently, e.g. looking down, turning around or swaying on the spot while waiting. 
Therefore, it is necessary to develop a method that captures only the motion in response to the external impulse.
In order to include all movements that are initiated by the external impulse, but at the same time exclude movements that do not relate to it, multiple criteria have to be fulfilled.
We investigate the forward velocities of the three body parts C7, Hip and the CoM and the acceleration of the CoM with equation \ref{eq:vf}.

We want to find the time $t_0$ at which the velocity of the participants is still zero.
But since only motions should be counted that exceed a threshold value, prospective events will be firstly investigated.
We assume that the CoM is considerably accelerated by the impulse, i.e. it is larger than $0.3\, \text{ms}^{-2}$.
If shortly afterwards the velocities of all three body parts C7, Hip and the CoM are greater than $0.05\, \text{ms}^{-1}$ simultaneously, it will be counted as motion due to the push.
This results in the Boolean variable $\text{motion}[i,t]$ that is true if the following condition occurs.

\begin{equation*} 
    \text{motion}[i,t]:
   a_i^{\text{CoM}}(t_1) > 0.3\, \frac{\text{m}}{\text{s}^2} \land  
   v_i^{\text{CoM}}(t_2)> 0.05\, \frac{\text{m}}{\text{s}} \land 
   v_i^{\text{C7}}(t_2) > 0.05\, \frac{\text{m}}{\text{s}}\land 
   v_i^{\text{HIP}}(t_2) > 0.05\, \frac{\text{m}}{\text{s}}, \quad \text{for } t_2 > t_1
\end{equation*}

The thresholds defining $\text{motion}[i,t]$ exclude minor movements that are not initiated by the external impulse. 
For the exact definition of the start of a phase, it is important to identify the point in time at which the participants are still at rest but will move in the next time step.
The time detected by $\text{motion}[i,t]$ deviates from the actual start time of motion.
Therefore, a backward search is performed to find the previous point in time at which the participant has not yet moved.
A participant stands definitely at rest, when the acceleration of the CoM is less than $0.15\, \text{ms}^{-2}$.
Using this threshold value, the start time $t_{\text{start}}[i]$ can be found as follows:

\begin{equation*} 
t_{\text{start}}[i] = t_0, \quad \text{if } a_i^{\text{CoM}}(t_0) > 0.15\, \frac{\text{m}}{\text{s}^2}, \quad \text{for } t_0 < t_2\\ 
\end{equation*}

The determined threshold values were found by comparing the output of the analysis with the side-view videos qualitatively to best represent the experiments (see Supplementary Information for more detail).
Figure \ref{fig:start_motion}\,a shows the forward acceleration, the threshold value of $0.3\, \text{ms}^{-2}$ as well as the start time $t_{\text{start}}[i]$.
The forward velocities of C7, Hip and the CoM are plotted as an example in Figure \ref{fig:start_motion}\,b.
The threshold value of $0.05\,\text{ms}^{-1}$ is shown as dotted grey line and the detected start time as dotted black line. 
In this example, the following times were detected: $t_1 = 0.68\,\text{s}$, $t_2 = 0.73\,\text{s}$ and $t_0 = t_{\text{start}}[i] = 0.67\,\text{s}$.

 \begin{figure}[H]
 \centering
\includegraphics[width=0.7\textwidth]{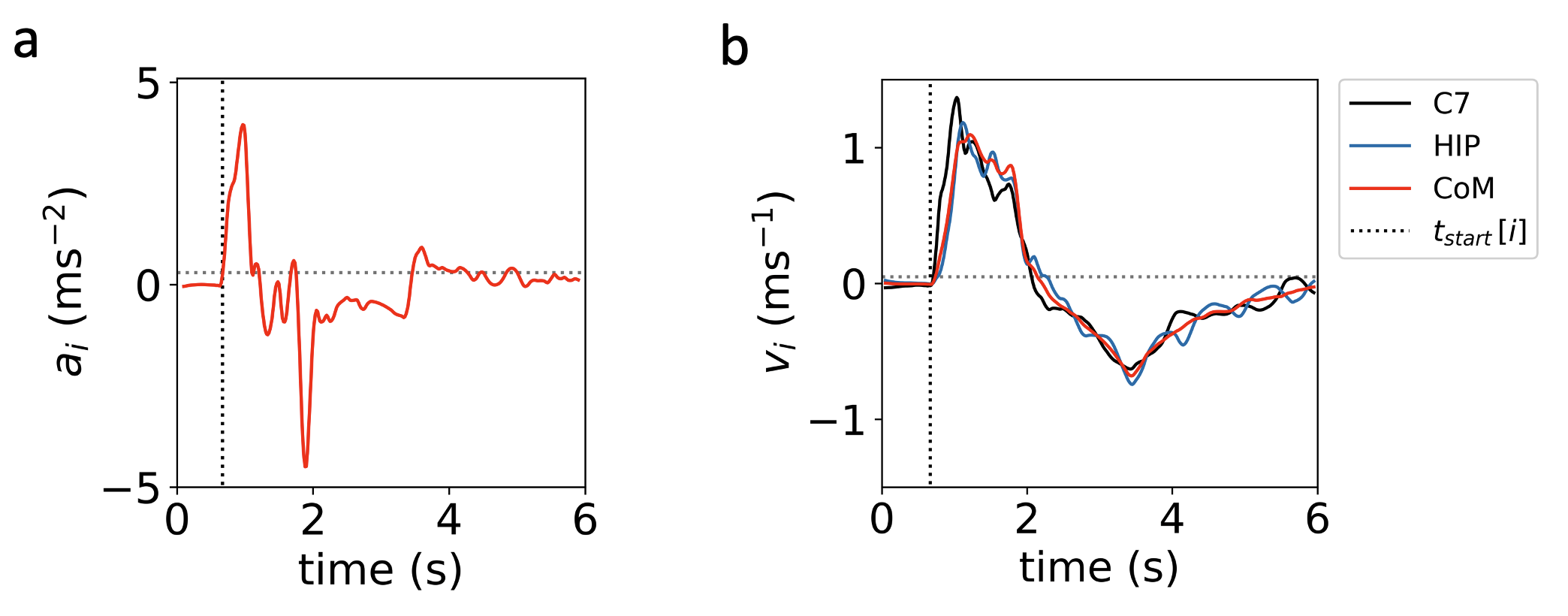}
\caption{(a) The forward acceleration of the CoM with a threshold value of $0.3\,\text{ms}^{-2}$. (b) Example of the velocity of C7, HIP and CoM with the threshold value of $0.05\,\text{ms}^{-1}$. The start time of the motion $t_{\text{start}}[i]$ is shown as dotted black vertical line.}
\label{fig:start_motion}
\end{figure}

\subsubsection{Perturbation and loss of standing balance}
\label{sec:perturbation}

As response to the external impulse, the participants are displaced from their rest position and potentially lose balance.
One way to measure standing balance is to consider the position of the projection of the CoM to the xy plane in relation to the position of the feet. 
The external points of both feet create the base of support (BoS).
In general static conditions, a person is considered in balance when the CoM lies inside the BoS and out of balance when the CoM is located outside of the BoS.
However, the CoM of individuals does not remain static following an external perturbation. 
In such dynamic conditions, the velocity of the CoM should also be considered to study standing balance.
Therefore, the margin of stability $MoS$ as proposed by \cite{hof_condition_2005}, which is the shortest distance of the extrapolated CoM to the boundary of the BoS, is analysed.

The extrapolated centre of mass XCoM takes the velocity into account and can be calculated with the pendulum equation $\omega_0 =  \sqrt{g/l}$. Here, $g$ is the gravitational acceleration and $l$ the leg length of the participant.
\begin{equation*} 
\mathbf{r}_i^{\text{XCoM}} (t) =  \mathbf{r}_i^{\text{CoM}} (t) + \frac{ \mathbf{v}_i^{\text{CoM}} (t)}{\omega_0}
\end{equation*}

For the calculation of the $MoS_i(t)$, we only consider the forward boundary of the BoS, i.e. the line TOE that passes through the two toe points (LT and RT), because we mainly investigate the effect of the impulse to the front.
With equation \ref{eq:distance3}, $MoS_i(t)$ is calculated as follows:

\begin{equation*} 
MoS_i (t) =  d(P_{xy} \mathbf{r}_i^{\text{TOE}}(t) , P_{xy} \mathbf{r}_i^{\text{XCoM}}(t))
\end{equation*}

\begin{figure}[H]
 \centering
\includegraphics[width=0.6\textwidth]{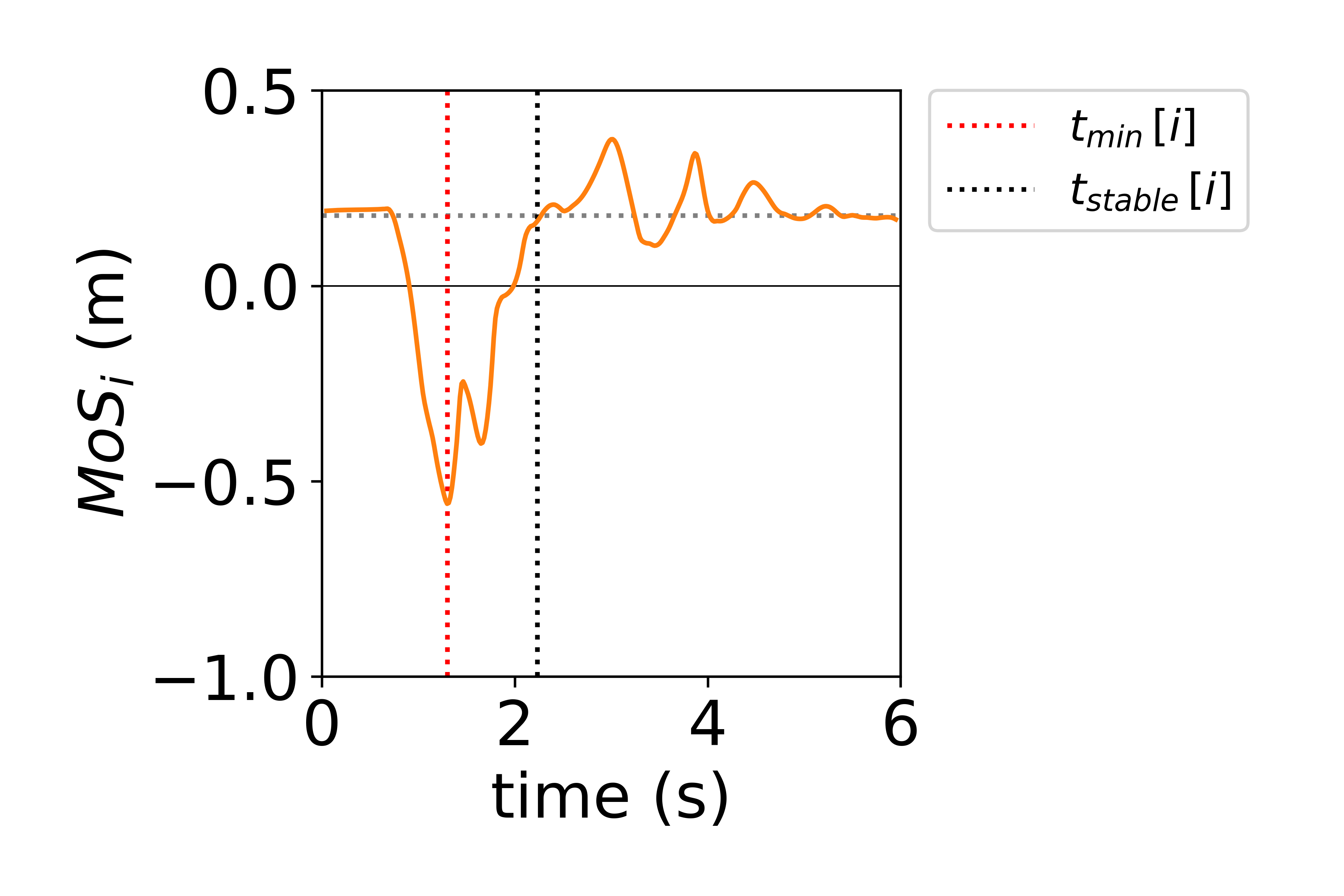}
\caption{Example of the $MoS$, which is the distance of the XCoM to the TOE line. The $MoS$ is positive when the XCoM lies ahead of the TOE line of the BoS and negative when lying behind the TOE line. The most unstable position at $t_{\text{min}}[i]$ is shown as dotted red vertical line and the time at which the person is in balance again $t_{\text{stable}}[i]$ is shown as dotted black vertical line.}
\label{fig:mos}
\end{figure}

This definition gives a negative distance when the XCoM lies ahead of the TOE line of the BoS and a positive distance when lying behind it.
An example of the $MoS_i(t)$ during one trial is shown in Figure \ref{fig:mos}.

Our goal is to distinguish between receiving and passing on the impulse.	
However, that is a very challenging task, because we have to differentiate between passive and active actions.
We assume that a participant is unintentionally displaced from their rest position and then takes action to regain a stable position.
Therefore, we can define the most critical state of standing balance corresponding to the smallest $MoS$.
This gives the time $t_{\text{min}}[i]$ until the person reaches the most unstable position.
After that, the person uses strategies to regain balance and thus may pass on the impulse.
To obtain an end time for potentially passing on the impulse, the actions of the participants need to be divided, in terms of what is a necessary reaction to the impulse and what is not. 
Since the $MoS$ also takes into account the velocity of the CoM as well as the positioning of the feet, we propose the time at which the $MoS$ reaches a threshold value as the end time. 
For our experiments, we define the threshold value as being 87\% of the initial $MoS_i$ value.
In the cases when this value is not reached, the time at which $MoS_i$ value is at a maximum within 2\,s after $t_{\text{min}}[i]$ is considered as end time.
However, these threshold values are just a starting point of the analysis that must be tested critically in further experiments (see Supplementary Information).

This results in the following definitions for both times. 

\begin{equation*} 
t_{\text{min}}[i] = \argmin_t MoS_i(t) 
\end{equation*}

\begin{equation*} 
t_{\text{stable}}[i] =
\begin{cases}
   \min[t_2], &\text{if } MoS_i (t_2)  \geq 0.87 \cdot MoS_i (0),  \text{ for } t_2 >  t_{\text{min}}[i] \\
    \argmax_{t_3} MoS_i(t_3), &\text{ for } t_{\text{min}}[i] < t_3 < t_{\text{min}}[i]  + 2\,\text{s}
\end{cases}
\end{equation*}\\

We estimate that at $t_{\text{stable}}$ the person has regained a stable position and no further motion is required to counteract the impulse.
Therefore, this is set as the time until the impulse could potentially be passed on, although this does not ensure whether this is actually the case. 
Additionally including the closest distance $d_{\text{min}}[i, i+1] $, as shown in Figure \ref{fig:d_time}, can give information on how close people are to each other and provide an indication of whether or not people are touching and thus passing on the impulse.
In order to find the time when people are out of contact again, we use the threshold as proposed in Section \ref{sec:touching}.
There is no longer a contact at time $t_{\text{touch}}$, when $d_{\text{min}} > 0.12\,\text{m}$.

\begin{equation*} 
t_{\text{touch}}[i] = \min[t_4], \quad \text{if } d_{\text{min}}[i, i+1](t_4) > 0.12\,\text{m} \quad \text{for } t_4 > t_{\text{start}}[i+1]
\end{equation*}

As a result, the end time $t_{\text{end}}$  of the last phase is given as the smallest time of $t_{\text{stable}}$ and $t_{\text{touch}}$.

\begin{equation*} 
t_{\text{end}}[i] = \min[ t_{\text{stable}}[i], t_{\text{touch}}[i] ]
\end{equation*}

\section{Results and Discussion}

\subsection{Detection of the phases}
Precisely detecting the defined phases of impulse propagation in the 3D MoCap data is a challenging task.
For this purpose, we used the following two methods, observation of the side-view videos as well as investigation of the 3D MoCap data.
We distinguished three phases, (0) initial position at rest, (i) receiving an impulse, (ii) receiving and passing on an impulse, and (iii) passing on an impulse.
In the videos, it is easier to see when people are touching each other rather than when each individual starts moving forward. 
For the 3D MoCap data, the opposite is the case: the start of motion is more reliable detected than the time people touch each other. 
However, touching does not necessarily involve passing on the impulse. 
Participants can grab the person in front instead of pushing forward, as sometimes seen in the videos.
Furthermore, people can move in a variety of ways and of their own choice and not every behaviour is a response to the impulse.
Therefore, it is important to evaluate which motions are counted for defining the phases and investigate various parameters simultaneously.\\

\begin{figure}[H]
 \centering
\includegraphics[width=0.65\textwidth]{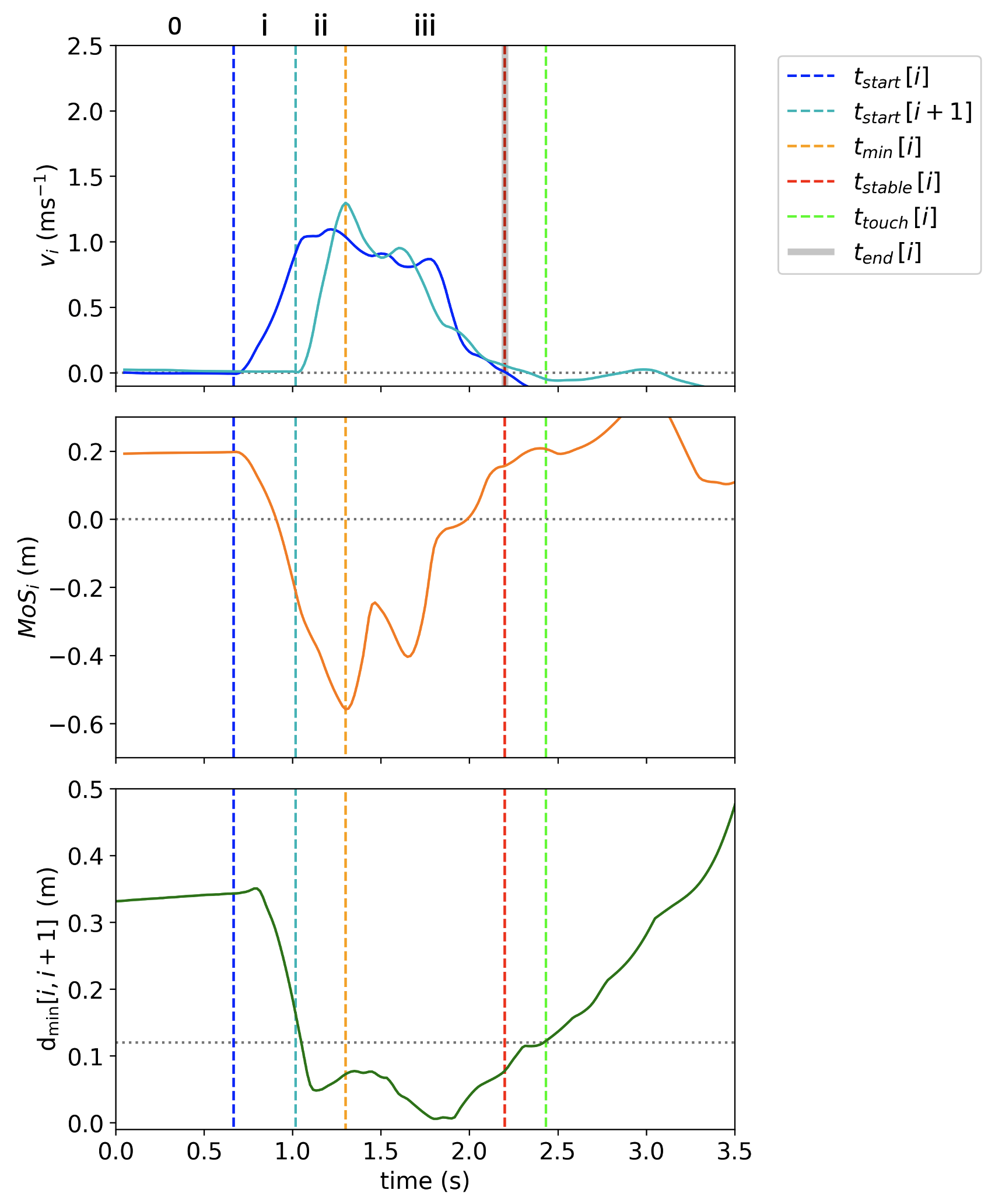}
\caption{Three variables are needed to identify the phases of motion for one person with the 3D MoCap data. The start time of the first phase is defined by the time the forward motion starts $t_{\text{start}}[i]$ and ends when the next person in the row starts moving forward $t_{\text{start}}[i+1]$. The second phase ends with the minimal $MoS$ at time $t_{\text{min}}[i]$. The end time of the third phase $t_{\text{end}}[i]$ is set by using the $MoS$ as well as the closest distance between two participants.}
\label{fig:summary}
\end{figure}

We come to the conclusion that mainly three variables are needed to identify the phases.
These are the velocity in the same direction as the external impulse, i.e. the forward direction, the margin of stability in the forward direction and the estimated distance between two participants.
Figure \ref{fig:summary} presents a summary of the variables as a function of time, and shows the start and end times of each phase for a trial without wall, in which the initial inter-person distance is set to elbow and the initial arm posture is free.
Based on this example, the following times are obtained for the first person pushed. 
Person $i=1$ receives the impulse and thus enters the first phase when they start moving forward ($t_{\text{start}}[1] = 0.67\,\text{s}$) . 
The second phase begins when the next person $i=2$ in the row moves ($t_{\text{start}}[2] = 1.02\,\text{s}$).
The start of the third phase $t_{\text{min}}[1] = 1.30\,\text{s}$ corresponds to the most unstable position, and it ends when the person reaches either a certain stability in the $MoS$ ($t_{\text{stable}}[1] = 2.20\,\text{s}$) or a certain distance to the person in front ($t_{\text{touch}}[1] = 2.43\,\text{s}$).
The phases continue smoothly into one another.
A general overview of these start and end times with the corresponding phases are listed in Table \ref{tab:phases2}.\\

\begin{table}[H]
\centering
    \caption{Overview of the phases of motion for each person in the middle of the row.}
    \label{tab:phases2}
    \begin{tabular}{l|ll|ll}
    	\toprule
        Phase & Impulse &  Behaviour & Start &  End \\
         \midrule
	0 & None  & In balance & $-$ & $t_{\text{start}}[i]$ \\
	&&&& \\
	i & Receiving & Moving forward, & $t_{\text{start}}[i]$ & $t_{\text{start}}[i+1]$\\
	 & &Displacement from rest position \\
   &&&& \\
	ii & Receiving and & Moving forward, Person in front moves, & $t_{\text{start}}[i+1]$ & $t_{\text{min}}[i]$\\
	& passing on & Displacement from rest position\\
  &&&& \\
	iii & Passing on & Person in front moves,  & $t_{\text{min}}[i]$ & $t_{\text{end}}[i]$ \\
	&& Regaining rest position\\
 &&&& \\
	iv & None & In balance  & $t_{\text{end}}[i]$ & $-$ \\
    \end{tabular}
\end{table}

Determining a time at which a person receives the impulse is relatively clear and the detection works well.
The beginning of our experiments represents a stationary crowd for example in a waiting scenario.
This means that participants do not move that much and there is a correspondingly significant motion, when they receive the impulse.
In order to capture only the motion in response to the external impulse, small as well as slow motions are not considered.
This ensures that independent movements, such as swaying on the spot, are excluded, but it also implies that slight effects on the body due to the impact are not taken into account.
Since the three body parts CoM, Hip and C7 have to move forward simultaneously to be included, movements like raising hands or looking down are neglected, and the forward component of the velocity causes rotations or moving backwards to remain undetected.
When comparing the identified start of motion with the videos, only one person is falsely detected (false positive) in one trial because they show a very similar motion pattern to a pushed person.
Furthermore, there are four cases of false negative detections.
These occur mainly when the persons are already touching each other while waiting for the impulse.\\

On the contrary, it is difficult to identify when a person is no longer receiving or passing on an impulse.
One can argue that touching as observed in the videos is an indicator, but on the other hand, this is no guarantee that forces are actually exchanged.
For example, one person could be holding the person in front of them instead of passing on the external impulse by pushing that person forward.
Therefore, we choose to use a measure of stability for the definition.
We are confident that up to the lowest $MoS$ ($t_{\text{min}}[i]$), which corresponds to a most unstable position, the impulse is received.
However, this does not necessarily mean that the person cannot continue to be affected by the impulse past this moment.
Since $t_{\text{min}}[i]$ sets the end of receiving in our definition, we may underestimate phase 2 and overestimate phase 3.
The experiments are designed in such a way that the subjects move forward due to the impulse, regain their balance and usually return to their starting position immediately afterwards. 
This returning is not caused by the impulse and should not be included as part of the phases.
Detecting the time $t_{\text{stable}}[i]$ when a person regained stability works quite well.
In our definition, we assume that no further action is necessary at this time to achieve a stable standing position, other than maybe placing the heels of the foot fully on the ground.
This outcome also agrees with the viewing of the videos, and only for a few single cases the time is specified a little late.\\

\subsection{Occurrence and duration of phases}

The three phases defined in Table \ref{tab:phases2} are not always applicable to all persons. 
By default, the last person to receive the impulse only goes through the first phase.
To investigate this, we examined the data for all four individuals within the row. 
The data for the furthermost person is ignored because, by definition, they can only undergo the first phase if at all and cannot pass on the impulse.
This means that a total of 420 pushed persons can be examined. 
In 28 cases, the person is not affected by the push when the initial inter-person distance is large (either arm or elbow) and therefore does not go through any phase. 
A further 28 people were the last person to be pushed and hence only experienced phase 1.
Of the remaining 364 pushed persons, 306 persons completed all three phases.

It could happen that there are sometimes just two phases.
If people touch each other directly at the beginning, which is particularly the case with arms up, but can also occur at no distance, people are able to transfer the impulse faster and thus shorten the time of the first phase.
For 7 persons, only phases 2 and 3 are detected and in extreme cases, only phase 2 is observed (9 persons). 
This means the difference of the start time of both participants  $t_{\text{start}}[i]$ and $t_{\text{start}}[i+1]$ is smaller than the time resolution of the sensors. Consequently, there are only phase 2 (receiving and passing on) and phase 3 (passing on the impulse).

By contrast, standing further apart at the beginning increases the probability that a person is already in the process of regaining a stable position by the time they reach the next person.
At large inter-person distances (elbow or arm), phase 2 is omitted (32 times) and therefore, only phases 1 and 3 are detected, which results in the phases (i) receiving and (iii) passing on the impulse.
In this scenario, there could be a gap between phases when $t_{\text{min}}[i] < t_{\text{start}}[i+1]$.
For a total of 10 persons, only phases 1 and 2 were identified, because the closest distance $d_{\text{min}}[i, i+1] $ never drops below the threshold of 0.12\,m.
\\

\begin{figure}[H]
 \centering
\includegraphics[width=0.65\textwidth]{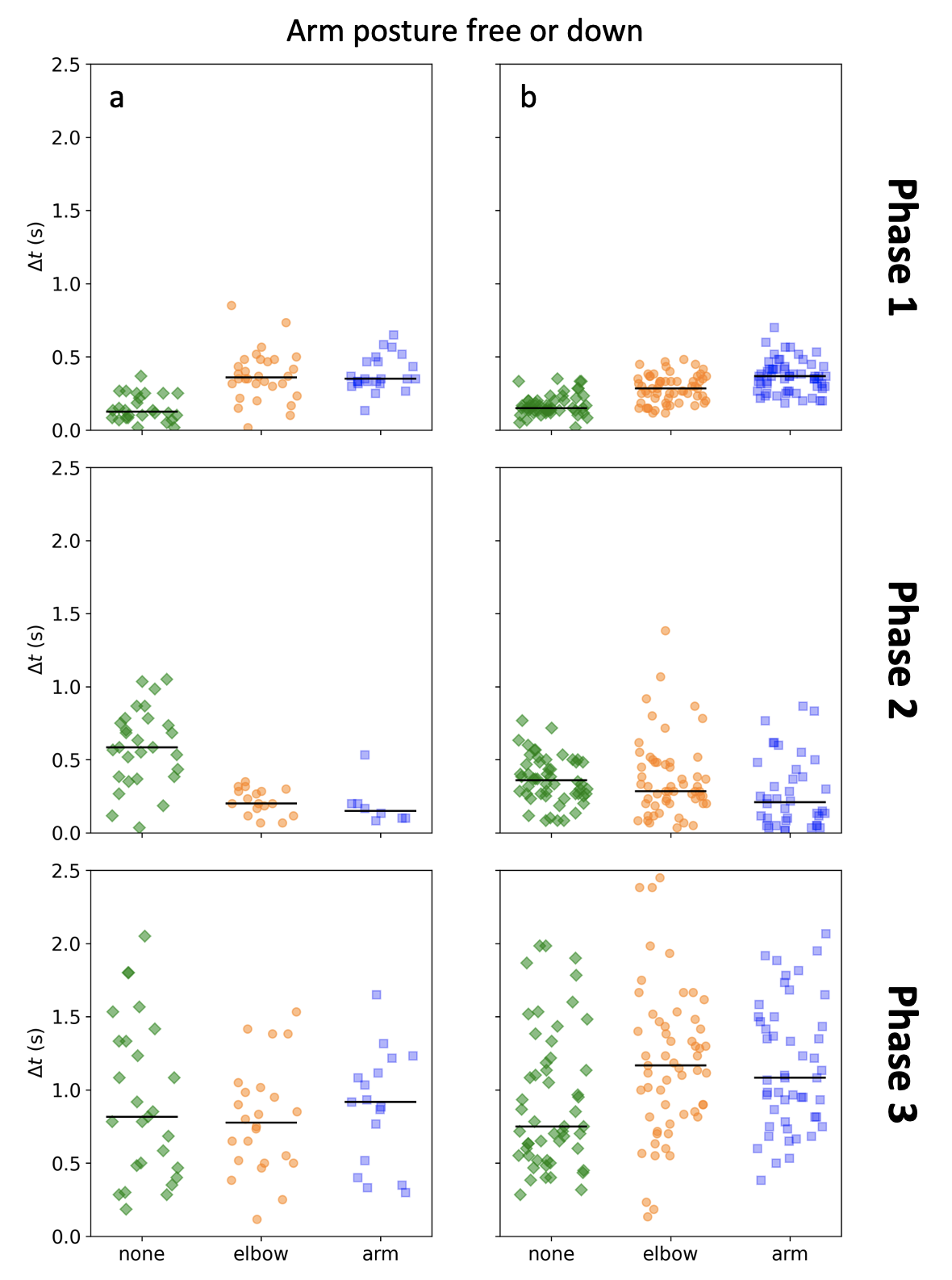}
\caption{Duration $\Delta t$ of phases 1, 2 and 3 according to different initial inter-person distances none, elbow and arm. The arm posture of participants were either free or down for trials (a) without a wall and (b) with wall. The median is shown as black horizontal line.}
\label{fig:result_d_NUp}
\end{figure}

\begin{figure}[H]
 \centering
\includegraphics[width=0.65\textwidth]{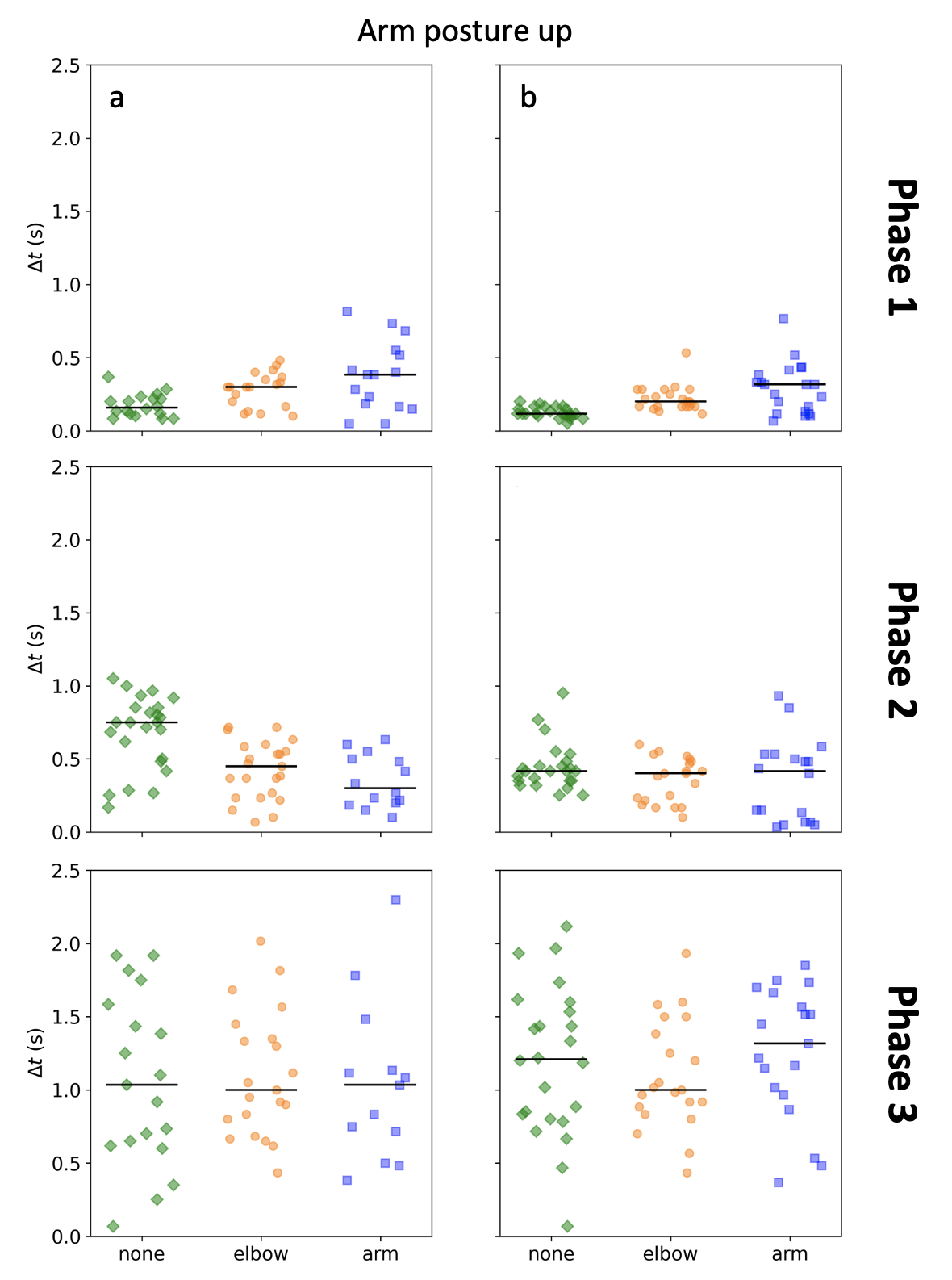}
\caption{Duration $\Delta t$ of phases 1, 2 and 3 according to different initial inter-person distances none, elbow and arm. The arm posture of participants were up for trials (a) without a wall and (b) with wall. The median is shown as black horizontal line.}
\label{fig:result_d_Up}
\end{figure}

In the next step of the analysis, the duration of each phase is investigated in relation to the initial distances and the arm position up or not up, whereby we distinguish between experiments with and without a wall. 
Figure \ref{fig:result_d_NUp} shows the duration of the phases $\Delta t$ for trials where the initial arm posture were either down or free and Figure \ref{fig:result_d_Up} for trials with arms up.

We find that the duration of phases 1 and 2 correlate with the initial inter-person distance which corresponds to our assumption. 
The closer the people stand to each other, the shorter is the first phase and the longer the second phase, as people interact with each other more quickly and therefore pass on the impulse sooner.
In addition, the options for motion strategies in phase 1 increase with the initial inter-person distance.
This is shown by the fact that the variance in phase duration is low for short distances and greater for longer distances.
No clear statement can be made about the third phase, because the duration is widely distributed across all intervals. 
In Phase 3, strategies to regain balance vary, and this variation is reflected in the variance of the phase duration, regardless of the initial inter-person distance.
Another reason could be the fact that the end of the third phase is difficult to determine (both in the qualitative analysis of the videos and on the basis of the data) and thus hard to detect.

All trends in the duration of the phases are given for both arm posture conditions (up or not up) and there is no significant difference between experiments with and without wall.
Furthermore, we did not find a correlation between the duration of the phases to the intensity of the impulses. Figures of this analysis can be found in the Supplementary Information. 
\\

\subsection{Applicability}

First of all, the aim of our research is to investigate how external impulses propagate in a crowd. 
It is important to note that people behave differently from billiard balls (elastic impact) and that the displacement of an individual can occur in various ways. 
This means that different characteristics of impulse transfer and strategies to regain balance can occur in a crowd. 
To investigate this in more detail and identify different characteristics, we divide the impulse transfer between people into different temporal phases.
We hope that these findings will be helpful in developing a physical model for impulse propagation in a crowd.
\\

Furthermore, it is of great importance to understand and characterise risks in a crowd.
Previously, there are different concepts when talking about these risks.
For example, a single person could fall down (\cite{sieben_inside_2023}), several people could move in one direction within the crowd leading to density waves (\cite{bottinelli_emergent_2016}), or the interactions between multiple people are described with a domino model (\cite{wang_modeling_2019}).
The temporal segmentation of motion propagation into three phases can contribute to an understanding how these different types of movement can emerge and help to identify the conditions under which they occur.

Phase 1 corresponds primarily to the fact that only one person is severely affected by the impulse. 
Particularly in cases where phase 2 is skipped, i.e. when a person moves from phase 1 to phase 3, there could be a risk of a single person falling down.
In phase 2, a person is not only receiving the impulse but also passes it on.
This could fit well with a domino model, as at least three people are simultaneously affected by the impulse.
To this end, the duration of the second phase could be investigated in more detail.
If the times of the second phase of several people overlap, it is possible to determine how many people are in the critical phase of losing balance at the same time and to examine wave movements of the crowd. 
The third phase indicates the time in which people regain their balance. 
Identifying this time could enable different movement strategies to be determined and automatically detected.
In addition, applying this method to rows with more people could identify the conditions under which an impulse may be intensified or dampened, allowing a more accurate description of impulse propagation in crowds.
\\

\subsection{Limitations}
It is interesting to note that there are differences in the closest inter-person distances between trials with and without a wall, as shown in Figure \ref{fig:touching}.
For the trials without wall, all medians are below the determined $0.12\,\text{m}$ threshold, while for the trials with a wall, three out of the four medians exceed the threshold.
The reason for this could be that the wall may cause people to change their strategies to transfer more of the impulse to the ground instead of passing it on to the next person.
As a result, participants keep more distance to their neighbours in this situation.
The contact detection method proposed in this study has a few weaknesses and could be improved.
Especially to investigate whether hands enclose arms or touch the shoulders at the back, the x-y plane must be considered in addition to the y-z plane.
This is also necessary for experiments that are extended to two dimensions.
The approximation by segments in the x-y plane, for example for the shoulders, can be useful as well. 
It would also be interesting to compare different methods that represent the participants differently. 
These would be for example point cloud, line segments or the convex hull.\\

Another limitation of this study is the accuracy of the data.
If the 3D MoCap data are not well enough aligned with each other due to combining the data with the head trajectories, it results in a significant error especially when calculating the distance between two people.
To minimise this error for the forward motion, we have calculated the forward direction for each person individually using the hip vector and did not use the y-direction as standard for all. 
Even within one person, the 3D MoCap data can contain some errors, for example if sensors are displaced during trials or if the body measurements, on which the articulation of the skeleton are based, are taken imprecisely.
Furthermore, this study \cite{guo_accuracy_2017} comes to the conclusion that the BoS calculated with Xsens showed a high error.\\

\section{Conclusion and Outlook}

In this study, the 3D motion of the human body as response to an external impulse is separated into temporal phases. 
These phases can be detected based on the MoCap data by investigating the forward velocity, the margin of stability and the estimated distance between two participants.
A maximum of three phases can occur, which are characterised by receiving, receiving and passing on or passing on the impulse.\\

Each phase should be further investigated in terms of their specific characteristics, physical interactions that occur, and the differences between the phases.
More research is also needed on the relation of the initial inter-person distance as well as the intensity of the impulse to the phases of motion
and the factors that influence them.
Possible factors include individual characteristics (e.g. height or weight of the participants), body tension, or also preparedness and reaction time.
The age of the participants should also be taken into account, as age may induce a number of changes in these characteristics.
As mentioned in \cite{tokur_review_2020} people choose different strategies, i.e. ankle, hip or steps, to regain a stable position.
As a result, the impulse might be transmitted differently or perhaps even be intercepted, leading to a general intensification or mitigation of the impulse along the row.
We could observe different movement strategies in the videos and it would be helpful to classify these strategies into individual movement types.
Thereby, the analysis could focus on pendulum movements of the upper body, the exact hip motions, the placement of the feet, and the use of the hands.\\

The experiments presented in this study could be enhanced to capture physical contacts by using additional pressure sensors on the hands and backs of the participants. 
This would also facilitate the estimation of normal forces acting between individuals and hence improve the analysis of whether an impulse is passed on. Furthermore, these experiments were conducted on a small scale corresponding to only a few real-life scenarios, such as queues.
Therefore, the analysis of the phases should be extended to larger experiments in which people are standing in multiple rows or groups.\\


\newpage%

\newpage

\renewcommand{\thetable}{S\arabic{table}}
\renewcommand{\thefigure}{S\arabic{figure}}
\setcounter{section}{0}

\section*{\large{Supplementary Information}}

 \setcounter{subsection}{0}
 \setcounter{table}{0}
  \setcounter{figure}{0}

\section{Analysed points of MoCap data}

\begin{table}[H]
\centering
    \caption{Overview of points of the human skeleton from the MoCap data that are analysed}
    \label{tab:MoCapData}
    \begin{tabular}{lrl}
    	\toprule
        \multicolumn{1}{c}{Description	} &  \multicolumn{2}{l}{c3d points} \\
         \midrule
         \multirow{2}{*}{Sternum} & \textbf{11} & pIJ \\
         & \textbf{14} & pPX \\
          \midrule
         \multirow{3}{*}{Hip} & \textbf{0} & pHipOrigin \\
         & \textbf{78} & jRightHip \\
         & \textbf{82} & jLeftHip \\
          \midrule
         \multirow{2}{*}{Spine} & \textbf{15} & pC7SpinalProcess \\
         & \textbf{8} & pL5SpinalProcess \\
         Sacrum & \textbf{7} & pSacrum \\
          \midrule
         \multirow{2}{*}{Hands} & \textbf{34} & pRightBallHand \\
          & \textbf{37} & pLeftBallHand \\
           \midrule
         \multirow{6}{*}{Feet} & \textbf{52} & pRightHeelFoot \\
         & \textbf{54} & pRightFifthMetatarsal \\
         & \textbf{57} & pRightToe \\
         & \textbf{63} & pLeftToe \\
         & \textbf{60} & pLeftFifthMetatarsal \\
         & \textbf{58} & pLeftHeelFoot \\
          \midrule
         CoM& \textbf{86} & CenterOfMass \\
    \end{tabular}
\end{table}

\section{Identification of the threshold values}

\subsection{Closest distance between participants}

\begin{figure}[H]
 \centering
\includegraphics[width=0.65\textwidth]{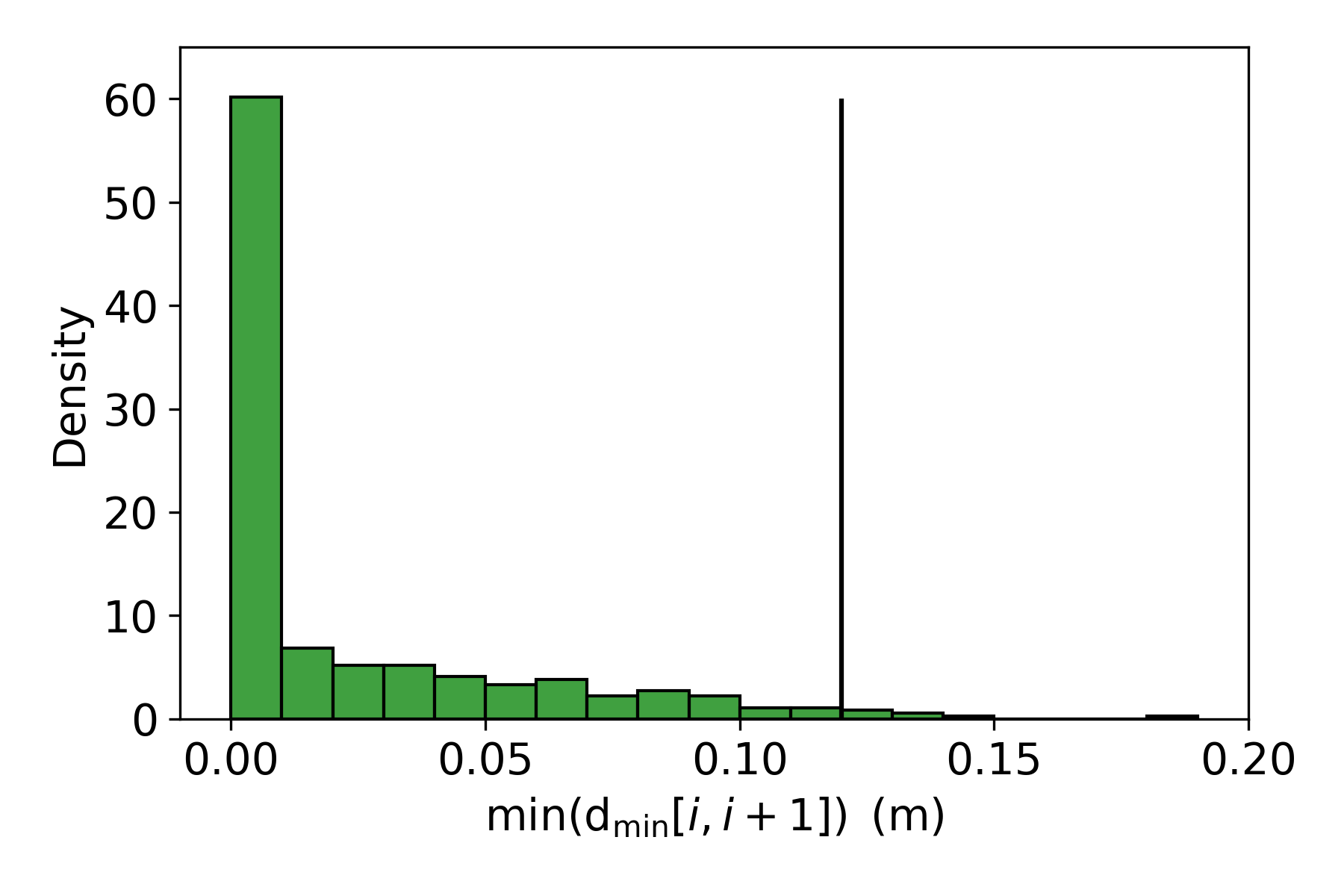}
\caption{Distribution of the global minima of the closest distances ($\min(d_{\text{min}}[i, i+1] )$) if the two participants $i$ and $i+1$ have moved forward. The threshold value of 0.12\,m is indicated as black vertical line.}
\label{fig:thr_dmin}
\end{figure}

In general, an impulse is passed from one person to the next when they touch each other. This observation can also be seen in the video recordings of the experiments. 
At this moment of contact, the closest distance between these two participants should theoretically be zero.
In practice, however, there is no guarantee that a calculated distance of zero will be reached, as we are only looking at skeletal data in which the extension of the body is neglected.
In addition, the calculated distances are superimposed by measurement errors and inaccuracies in the data fusion.
Nevertheless, we assume that the smallest distance between the participants will reach a minimum value over time at which a contact is very likely. 
In order to determine a threshold value, we consider the minimum of the closest distance ($\min(d_{\text{min}}[i, i+1] )$) of all trials in which both participants $i$ and $i+1$ have moved forward (see Figure \ref{fig:thr_dmin}). These trials represent the correct positive cases of passing the motion.
It can be seen, that in most cases the smallest distance is close to zero. 
To include as many trials as possible and keep the threshold as small as possible, we choose the threshold of 0.12\,m, which is plotted as black vertical line. 
This threshold value retains 98\% of the cases.

\subsection{Acceleration and velocity}

\begin{figure}[H]
 \centering
\includegraphics[width=0.65\textwidth]{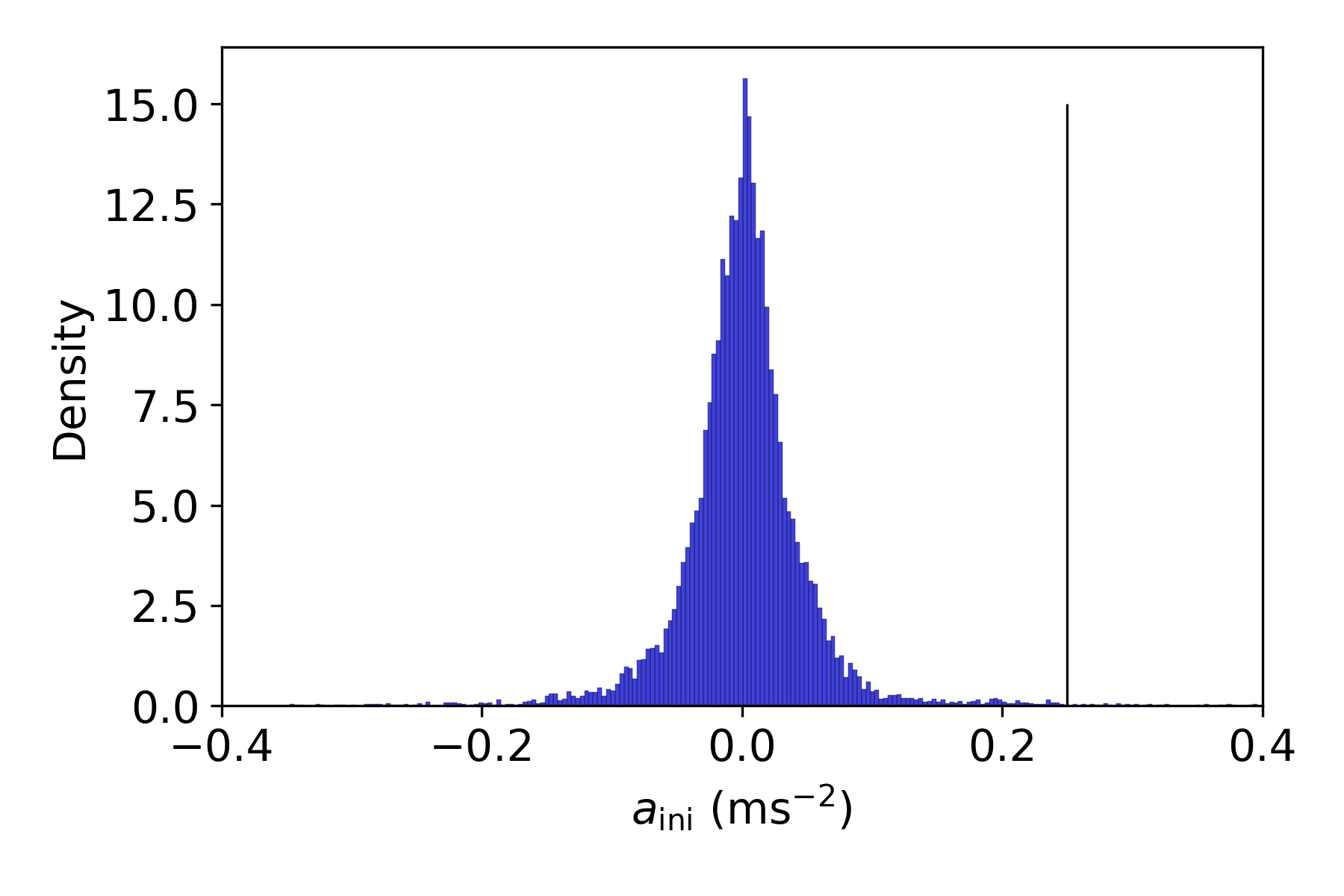}
\caption{Distributions of all measured accelerations in the first 0.5 seconds of a trial The participants were not yet pushed and were standing on the same spot.}
\label{fig:thr_aini}
\end{figure}

People can move individually and independently, so that velocities and accelerations are measured continuously, even if usually within a small range. 
However, larger values can also be achieved if, for example, participants look down at the floor, turn around or sway on the spot while waiting.
Since we assume that the participants are accelerated quite strongly by an external impulse, the acceleration should differ from the resting movements. 
In order to be able to better estimate how large values become in a stable waiting position, the accelerations of the CoM of the first 0.5\,s are examined for all trials, as there was no external impulse at that time.
Based on the distribution of these accelerations in Figure \ref{fig:thr_aini}, we assume that the acceleration should be greater than $0.25\,ms^{-2}$ in order to ensure a reliable and at the same time not too sensitive detection of motion.
\\

\begin{figure}[H]
 \centering
\includegraphics[width=0.65\textwidth]{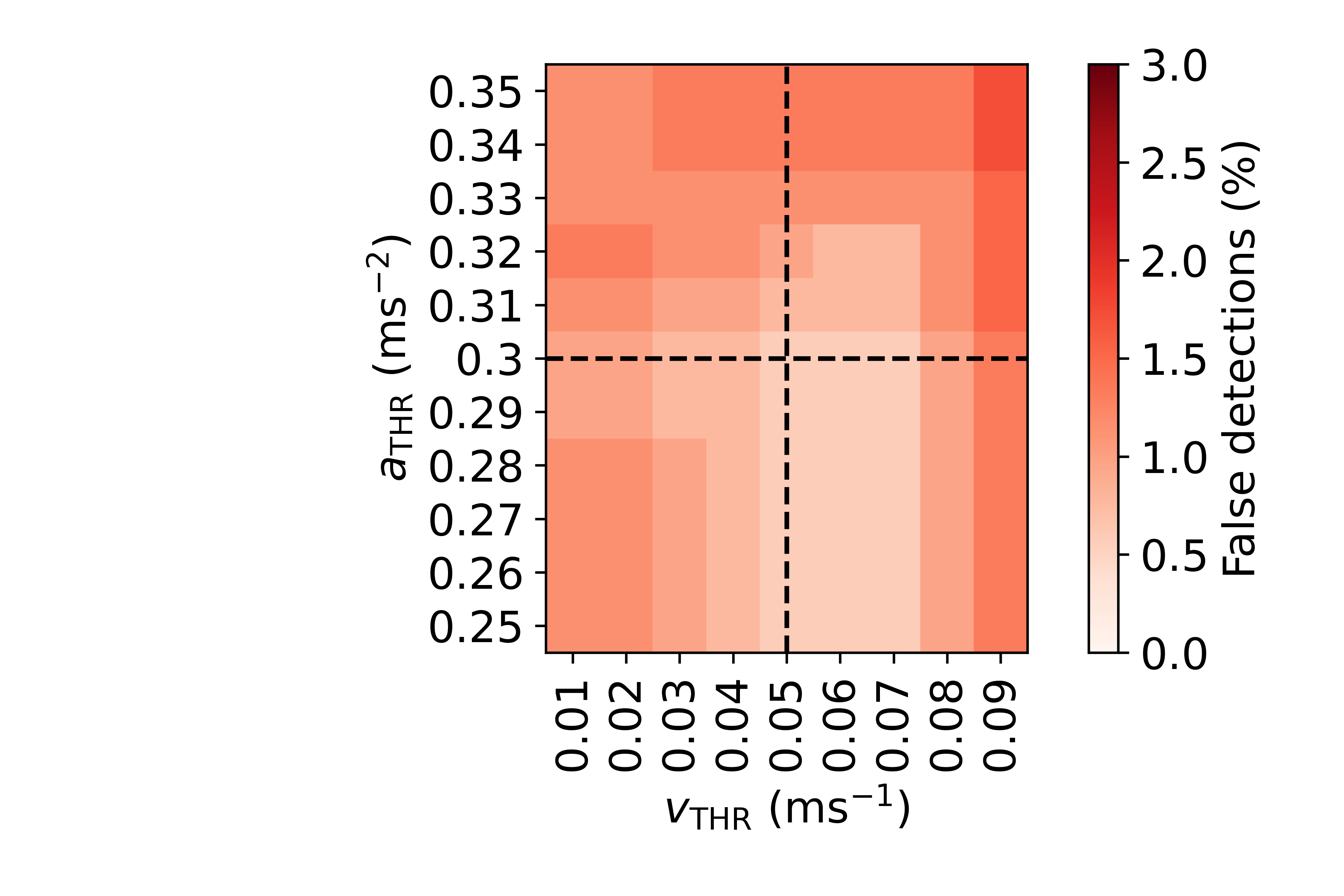}
\caption{False detections for different threshold values for the acceleration $a_\text{THR}$ and the velocity $v_\text{THR}$ when compared to the side-view videos. For further analysis, we choose the combination of the highest acceleration ($a_\text{THR} = 0.3\,ms^{-2}$) and the smallest velocity ($v_\text{THR} = 0.05\,ms^{-1}$) which has the best detection rate.}
\label{fig:thr_av}
\end{figure}

In a next step, we compare different threshold values of the acceleration and the velocity with the side-view videos.
The videos observation do not provide a ground truth in the sense of a quantitative measurement, but rather a qualitative yes or no observation. 
From watching the videos, the start times of motion can not be precisely determined.
Instead, the number of participants that are affected by the impulse can be easily and quickly identified.
Furthermore, this value can be compared to the Boolean variable motion[i, t] from section \ref{sec:motion} to estimate false positive and false negative detections.
Various threshold values for the acceleration $a_\text{THR}$ and the velocity $v_\text{THR}$ are tested and all false detections are summed up without distinguishing between false negative and false positive.
Theses false detections are shown in red in Figure \ref{fig:thr_av} indicating a more accurate detection in a lighter colour.
For further analysis, the highest acceleration and the lowest velocity with the best detection rate are choosen resulting in threshold values of  $a_\text{THR} = 0.3\,ms^{-2}$ and $v_\text{THR} = 0.05\,ms^{-1}$.
The latter threshold value $v_\text{THR} = 0.05\,ms^{-1}$ is in accordance with the adopted threshold in perturbation experiments of single persons (\cite{chatagnon_stepping_2023}).

\subsection{Perturbation and loss of standing balance}

The search for the end of phase 3 turned out to be very challenging. 
During this phase, the participants try to regain their balance by using different strategies, such as taking steps forward, leaning on the person in front and straightening up again, etc..
This results in different movement patterns. 
Furthermore, participants usually returned to their starting position immediately after regaining their balance. 
It is therefore also necessary to differentiate which movements were made in response to the impulse and which movements are no longer absolutely necessary, which makes it even more difficult. 
The challenge here is to find a coherent  parameter that includes as many strategies as possible, but at the same time is not too large to find a suitable ending.
In addition, losing and regaining balance cannot be precisely recognised in the videos.
In our investigation, no set of external characteristics could be found to determine this in a standardised way.

We have decided to investigate the $MoS$, as not only the position of the CoM in relation to the feet but also the velocity of the CoM is taken into account. 
Thus, a certain value of $MoS$ can only be achieved when the person is upright and the feet and CoM move slightly.
Moreover, it is a measure that is then calculated the same for everyone and simplifies a comparison.
A maximum value of the $MoS$ is not very sufficient, as participants can slowly regain balance and not always a local maximum is reached.
To simplify the analysis, we wanted to use the initial value of the $MoS$ as the threshold value, because we assume that the chosen starting position of each participant is a stable standing position.
However, people do not necessarily have to return to their original state, so this value is not always reached after having received the push.
On the other hand, the threshold value should not be set too low, as the displacement was sometimes small.
From an observational analysis, the 87\% of the $MoS$ was chosen to include as most participants as possible. 
Alternatively, 2\,s  were selected to give participants enough time (typical reaction time of young adults ranges from 0.18 s to 0.3 s (\cite{Peon-2013,Aditya-2015}) to regain their balance and determine an end.
This threshold is just a starting point of the analysis that must be tested critically in further experiments.

\section{Analysis of the duration of phases relative to the intensity of external impulses}

The duration of the three phases is investigated in relation to the intensity of the external impulses and differentiated between experimental conditions.
No significant correlation in terms of intensity was found for any of the phases. 

\begin{figure}[H]
 \centering
\includegraphics[width=0.65\textwidth]{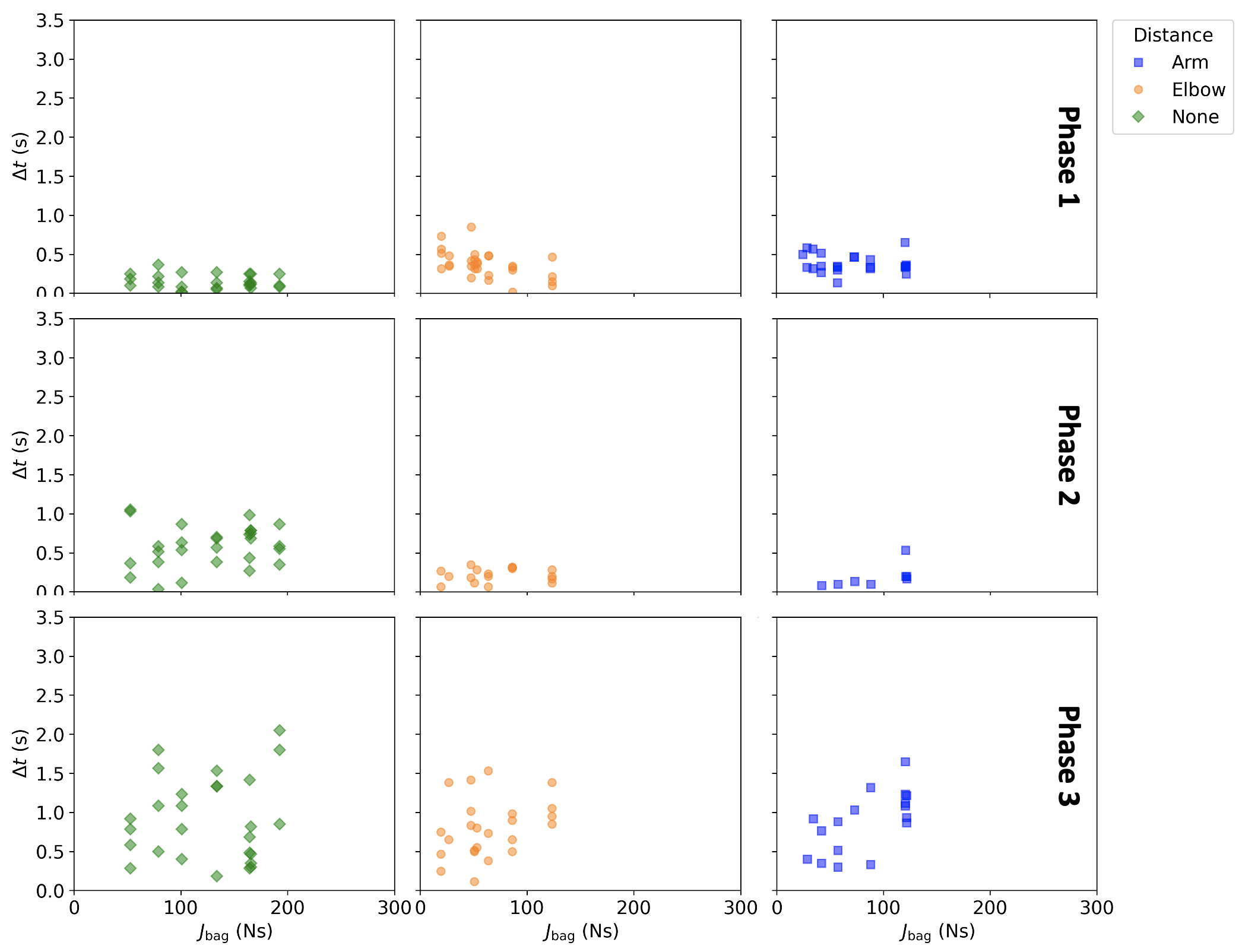}
\caption{Duration $\Delta t$ of phases 1, 2 and 3 in relation to the intensity of the impulses for the different initial inter-person distances none, elbow and arm. The arm posture of participants were either free or down for trials without a wall.}
\label{fig:J_NUp_noW}
\end{figure}

\begin{figure}[H]
 \centering
\includegraphics[width=0.65\textwidth]{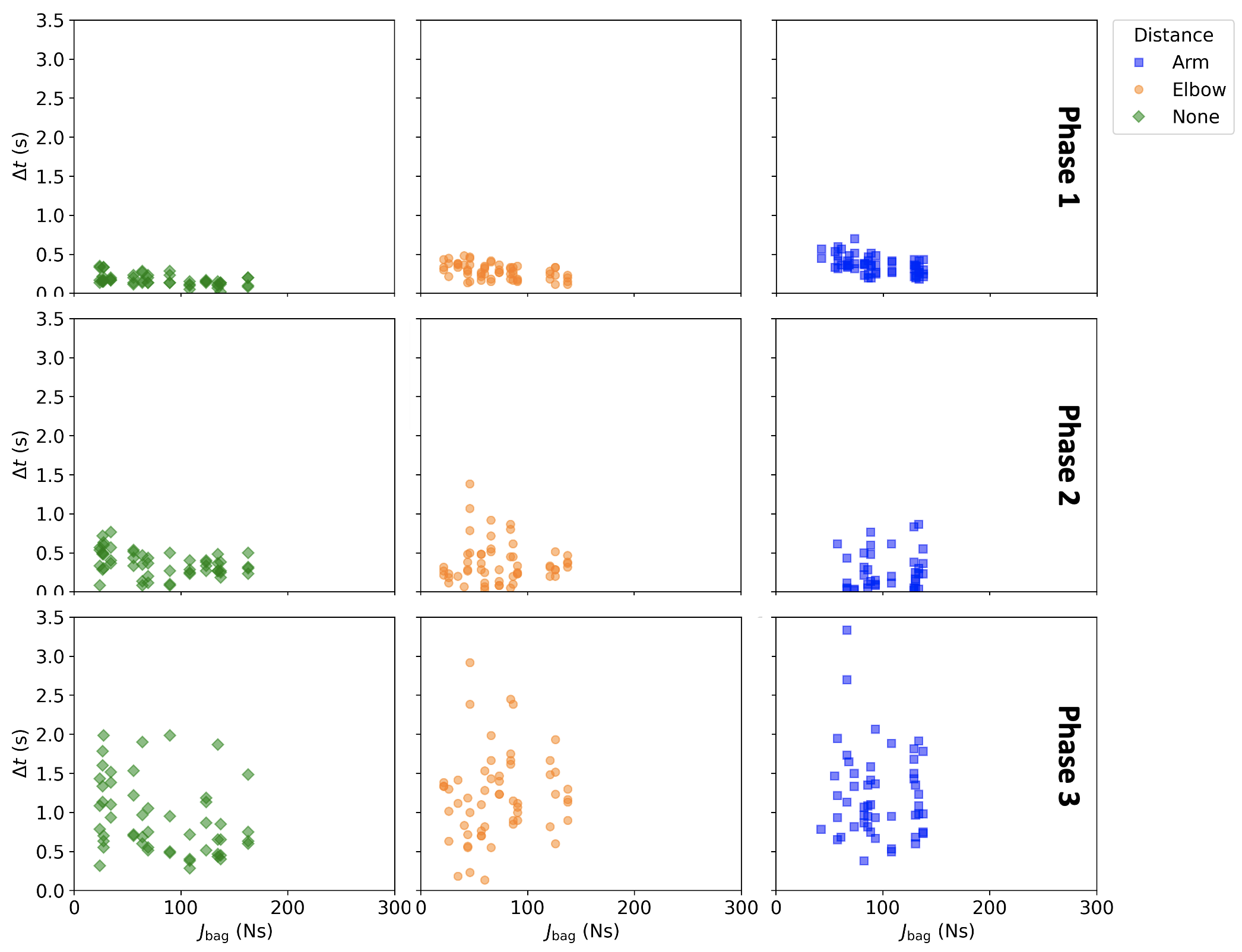}
\caption{Duration $\Delta t$ of phases 1, 2 and 3 in relation to the intensity of the impulses for the different initial inter-person distances none, elbow and arm. The arm posture of participants were either free or down for trials with a wall.}
\label{fig:J_NUp_wiW}
\end{figure}

\begin{figure}[H]
 \centering
\includegraphics[width=0.65\textwidth]{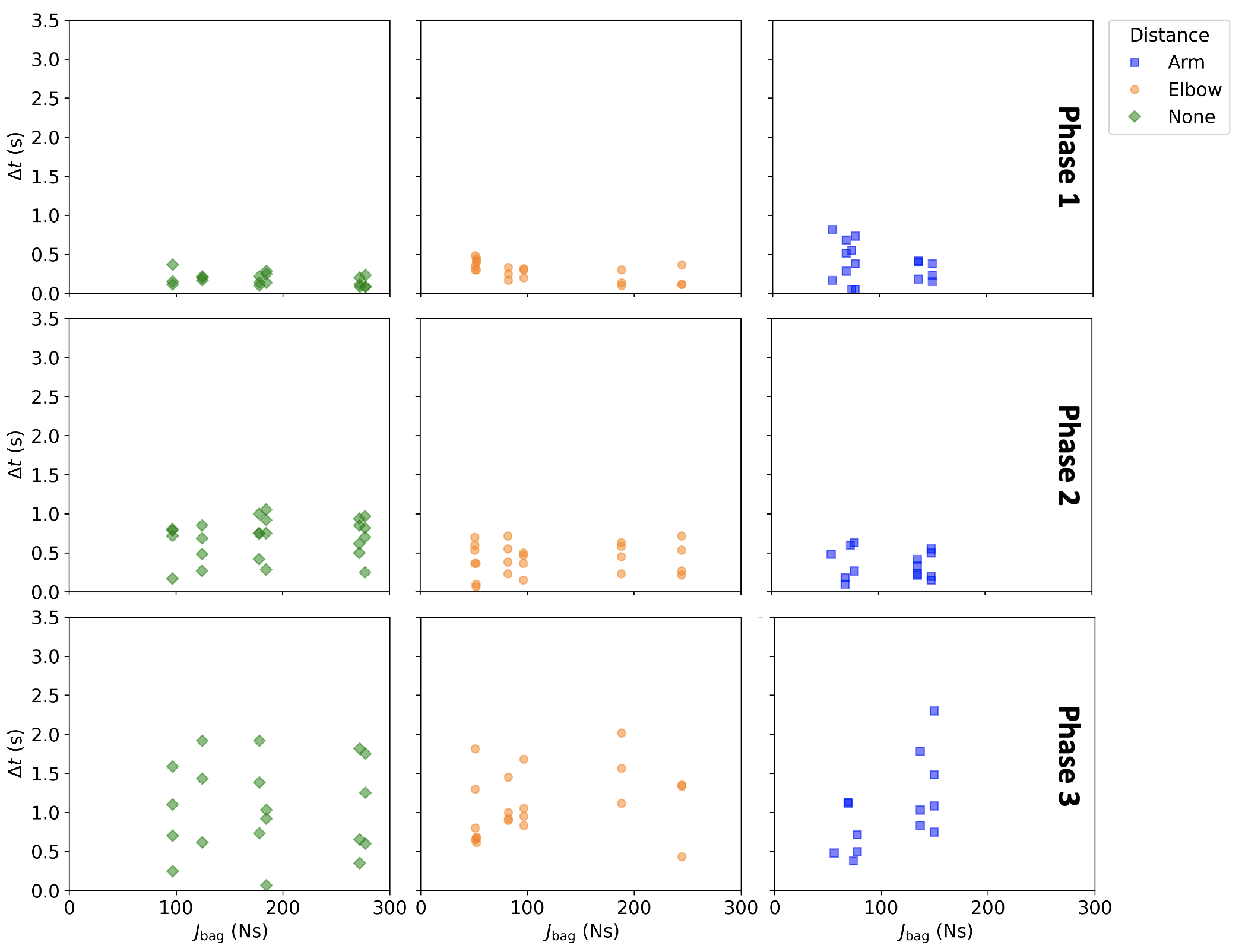}
\caption{Duration $\Delta t$ of phases 1, 2 and 3 in relation to the intensity of the impulses for the different initial inter-person distances none, elbow and arm. The arm posture of participants were up for trials without a wall.}
\label{fig:J_Up_noW}
\end{figure}

\begin{figure}[H]
 \centering
\includegraphics[width=0.65\textwidth]{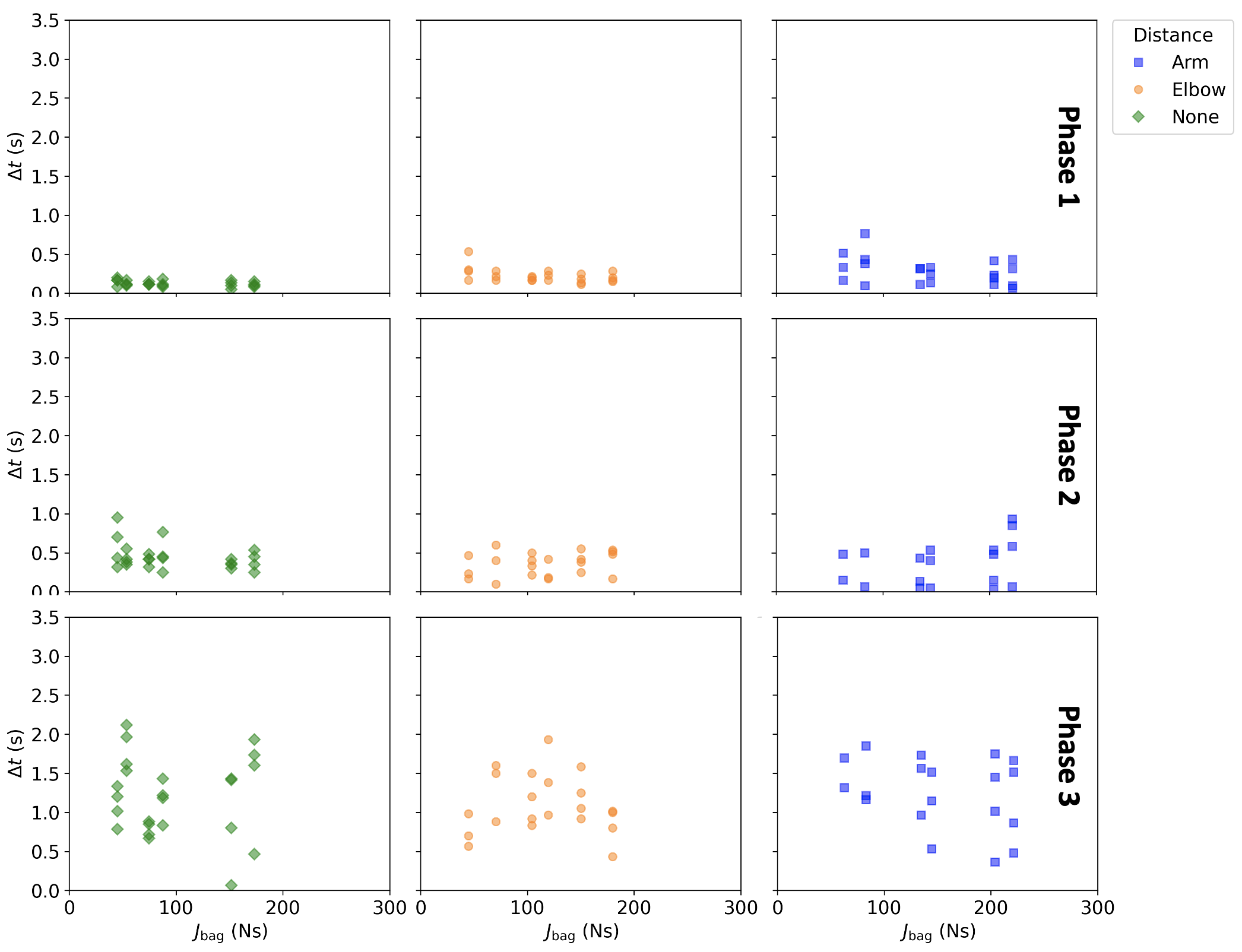}
\caption{Duration $\Delta t$ of phases 1, 2 and 3 in relation to the intensity of the impulses for the different initial inter-person distances none, elbow and arm. The arm posture of participants were up for trials with a wall.}
\label{fig:J_Up_wiW}
\end{figure}

\end{document}